\documentclass[journal=jctcce, manuscript=article, layout=traditional]{achemso}

\usepackage[version=3]{mhchem} 
\usepackage[T1]{fontenc}       
\usepackage{color}
\usepackage{amsmath}
\usepackage{amsfonts}
\usepackage{graphicx}
\usepackage{verbatim}
\usepackage{hyperref}
\usepackage{braket}
\usepackage{algorithm}
\usepackage{algpseudocode}

\algdef{SE}[DOWHILE]{Do}{doWhile}{\algorithmicdo}[1]{\algorithmicwhile\ #1}

\title[Practical guide to DMET]{A practical guide to density matrix embedding theory in quantum chemistry}
\author{Sebastian {Wouters}}
\email{sebastianwouters [at] gmail.com}
\affiliation{Department of Chemistry, Princeton University, Frick Chemistry Laboratory, Princeton, NJ 08544, USA}
\author{Carlos A. {Jim\'enez-Hoyos}}
\affiliation{Department of Chemistry, Princeton University, Frick Chemistry Laboratory, Princeton, NJ 08544, USA}
\author{Qiming {Sun}}
\affiliation{Department of Chemistry, Princeton University, Frick Chemistry Laboratory, Princeton, NJ 08544, USA}
\author{Garnet K.-L. {Chan}}
\email{gkc1000 [at] gmail.com}
\affiliation{Department of Chemistry, Princeton University, Frick Chemistry Laboratory, Princeton, NJ 08544, USA}

\begin{document}

\begin{abstract}
Density matrix embedding theory (DMET)
[\href{http://dx.doi.org/10.1103/PhysRevLett.109.186404}{Knizia and
    Chan, \textit{Phys. Rev. Lett.} \textbf{109}, 186404 (2012)}]
provides a theoretical framework to treat finite fragments in the
presence of a surrounding molecular or bulk environment, even when
there is significant correlation or entanglement between the two.  In
this work, we give a practically oriented and explicit description of
the numerical and theoretical formulation of DMET. We also describe in
detail how to perform self-consistent DMET optimizations. We explore
different embedding strategies with and without a self-consistency
condition in hydrogen rings, beryllium rings, and a sample
S$_{\text{N}}$2 reaction.  The source code for the calculations in
this work can be obtained from
\url{https://github.com/sebwouters/qc-dmet}.
\end{abstract}

\begin{tocentry}
\centering
\includegraphics[width=3.52in,height=1.37in]{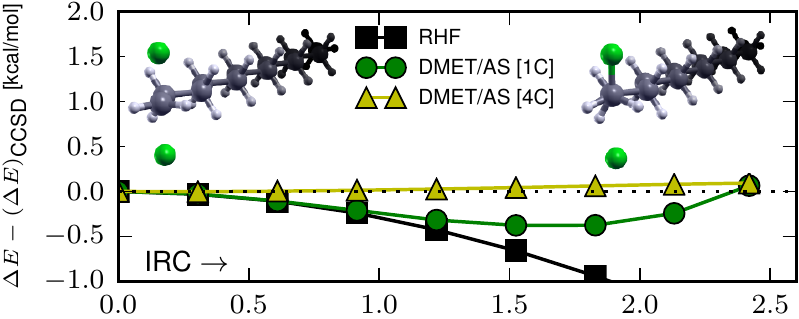}
\end{tocentry}

\maketitle
\section{Introduction}

Many quantum systems require a treatment beyond mean-field theory to
adequately capture properties of interest. A long-standing problem in
quantum many-body theory has therefore been the development of
computationally feasible and accurate correlated methods. This problem
has been explored in the contexts of nuclear structure, condensed
matter, and quantum chemistry, quite often with significant
cross-fertilization. Methods such as coupled-cluster
theory,\cite{Coester1960477, CCcizek, CClattice} the density-matrix
renormalization group (DMRG),\cite{PhysRevLett.69.2863, whiteQC,
  PhysRevC.63.061303} and dynamical mean-field theory
(DMFT)\cite{PhysRevLett.62.324, PhysRevLett.69.1240, zgidDMFT,
  PhysRevLett.106.096402}, are examples of techniques now employed
across different branches of physics and chemistry.

Density matrix embedding theory (DMET) is another
example.\cite{geraldPRL,geraldJCTC} Its foundation lies on the border
between tensor network states (TNS) and DMFT. TNS provide a versatile
framework for reasoning about the quantum entanglement of local
fragments with their surrounding neighbours in terms of the Schmidt
decomposition of quantum many-body states,\cite{EPJDreview} while DMFT
self-consistently embeds the Green's function of local fragments in a
fluctuating environment.\cite{RevModPhys.68.13} While TNS are able to
capture the low-lying eigenstates to high accuracy, they require an
explicit representation of the {\it entire} quantum many-body system
at the same level of approximation; even with translational
invariance, accurate contractions of the environment have to be
performed. DMFT circumvents this problem by treating only the local
fragment at an explicit self-consistent many-body Green's function
level, with the environment represented only by its hybridization
function.  However, non-local interactions between the fragment and
its environment become more difficult to include in DMFT, and the
formulation in terms of frequency dependent quantities engenders
additional numerical effort over ordinary ground-state calculations.

DMET attempts to combine the best of the two worlds, and in doing so
introduces an approximation of its own. Similar to DMFT, DMET embeds a
local fragment, treated at a high level, in an environment, treated at
a low level, thus circumventing the need to represent the entire
system with uniform accuracy. However, in contrast to DMFT which
embeds the Green's function, the embedding of DMET uses only the
ground-state density matrix, and thus does not require a
frequency-dependent formulation. The accuracy of DMET depends on the
low-level and high-level methods that enter into the formulation. The
low-level method is used to provide an approximate ground-state
wavefunction, from which a bath space for the local fragment is
obtained by a Schmidt decomposition. The high-level method computes a
wavefunction in the space of the local fragment with the small number
of bath states, to high accuracy.  DMET is thus a kind of wavefunction
in wavefunction embedding method and there can be a rich variety of
combinations of low-level and high-level methods.  For example, some
low-level methods that have been used in DMET are Hartree-Fock (HF)
theory,\cite{geraldPRL, geraldJCTC} Hartree-Fock-Bogoliubov
theory,\cite{boxiaoHubbard} anti-symmetrized geminal power (AGP)
wavefunctions,\cite{troyGeminals} coherent state wavefunctions for
phonons,\cite{barbara} and block product states for
spins.\cite{spinsystemPRB} Some examples of high-level methods that
have been used are exact diagonalization (also known as full
configuration interaction (FCI)),\cite{geraldPRL, geraldJCTC}
DMRG,\cite{qiaoniPRB, boxiaoHubbard} and coupled-cluster
theory.\cite{bulikJCP}

So far, the ground-state formulation of DMET has been the most widely
applied. In condensed matter systems it has been used to study the
one-dimensional Hubbard model,\cite{geraldPRL, bulikPRB} the
one-dimensional Hubbard-Anderson model,\cite{troyGeminals} the
one-dimensional Hubbard-Holstein model, \cite{barbara} the
two-dimensional Hubbard model on the square\cite{geraldPRL,
  boxiaoHubbard, PhysRevX.5.041041} as well as the honeycomb
lattice,\cite{qiaoniPRB} and the two-dimensional spin-$\frac{1}{2}$
$J_1$-$J_2$-model.\cite{spinsystemPRB} Quantum chemistry applications
have been fewer, but it has been used to study hydrogen rings and
sheets,\cite{geraldJCTC} as well as carbon polymers, two-dimensional
boron-nitride sheets, and crystalline diamond.\cite{bulikJCP} We also
want to mention that the DMET bath orbital construction can be used to
define optimal QM/MM boundaries,\cite{qimingJCTC} as well as to
construct atomic basis set contractions which are adapted to their
chemical environment.\cite{geminal_ao_construction} While DMET has
mainly been used for ground-states, though, the formalism is not
limited to ground-state properties. By augmenting the ground-state
bath space with additional correlated many-body states from a Schmidt
decomposition of the response wavefunction, accurate spectral
functions have been obtained.\cite{georgePRB, qiaoniPRB}

Despite this growing body of work on DMET from several workers, our
own group's presentation of the numerical implementation and
theoretical formulation of DMET has been limited to the two short
original articles~\cite{geraldPRL, geraldJCTC} and the supplementary
information of Ref.~\citenum{boxiaoHubbard}. The discussion of our
implementation for quantum chemistry problems has been particularly
brief. This work therefore attempts to provide a more explicit
explanation of DMET from our perspective, that we believe will be
particularly useful for those seeking to implement the method for
their own chemistry applications. Together with this work, we provide
a code \textsc{qc-dmet}\citep{github_qcdmet} that may be used in real
calculations. For simplicity, we focus exclusively on the ground-state
formulation of DMET.

In Sec. \ref{section_bath_orbital_construction} we begin by discussing
the DMET bath construction. The DMET low-level and high-level
embedding Hamiltonians, and their connection through self-consistency,
are then introduced and their construction is explained in
Sec. \ref{active_space_ahm_major_Section}.  We explain how to compute
expectation values (such as the energy) from the one- and two-particle
reduced density matrices of the ground states of the embedding
Hamiltonians in different fragments in
Sec. \ref{DMETenergy_major_seciton}. The numerical aspects of the
self-consistency of DMET are treated in
Sec. \ref{correlationpotentialoptimization_subsection}. Various
algorithmic choices are tested, and their implications are discussed,
in Sec. \ref{section_applications}. In Sec. \ref{section_summary}, we
summarize our results.

\section{The DMET bath construction} \label{section_bath_orbital_construction}

\begin{figure}
 \includegraphics[width=0.2\textwidth]{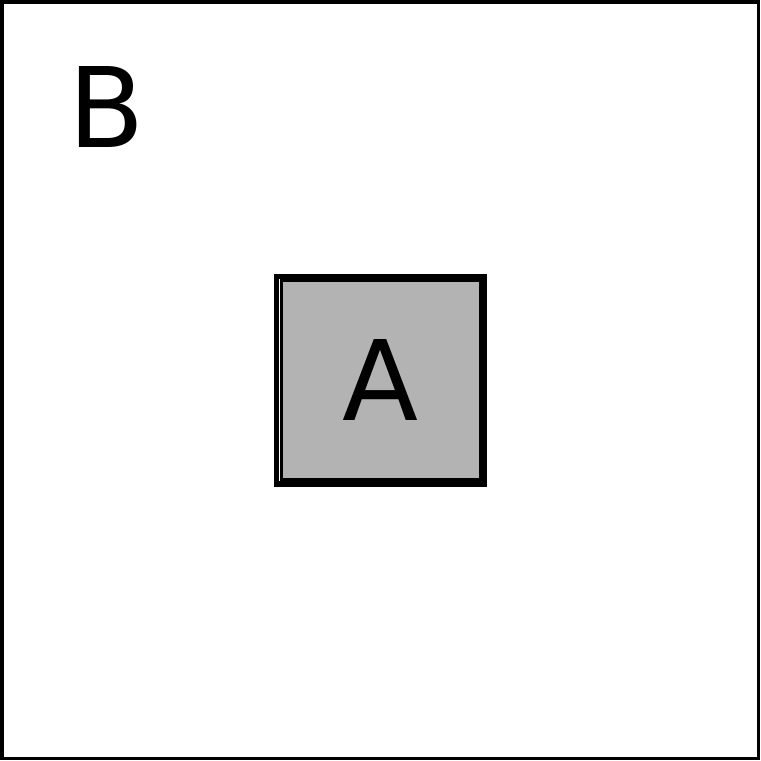}
 \caption{\label{splitupabfig} Local fragment A and its environment B.}
\end{figure}

Imagine a system composed of two parts, a fragment (typically called
an impurity in lattice applications) A and an environment B as shown
in Fig. \ref{splitupabfig}. In general, any wavefunction $\ket{\Psi}$
of the full system can be expressed in the Hilbert space of the states
of A and B, i.e. $\{ \ket{A_i} \otimes \ket{B_j}\}$, of dimension $N_A
\times N_B$. However, if the $\ket{\Psi}$ of interest is known a
priori, its Schmidt decomposition for the local fragment A and its
environment B allows to reduce the number of required many-body states
for the environment B significantly:
\begin{eqnarray}
\ket{\Psi} & = & \sum\limits_{i}^{N_A} \sum\limits_{j}^{N_B} \Psi_{ij} \ket{A_i} \ket{B_j} \nonumber \\
             & = & \sum\limits_{i}^{N_A} \sum\limits_{j}^{N_B} \sum\limits_{\alpha}^{\min(N_A,N_B)} U_{i\alpha} \lambda_{\alpha} V^{\dagger}_{\alpha j} \ket{A_i} \ket{B_j} = \sum\limits_{\alpha}^{\min(N_A,N_B)} \lambda_{\alpha} \ket{\widetilde{A}_{\alpha}} \ket{\widetilde{B}_{\alpha}}. \label{schmidtdecomp}
\end{eqnarray}
We remind the reader that the singular value decomposition of the
coefficient tensor $\Psi_{ij}$ yields two unitary basis
transformations $U_{i\alpha}$ and $V_{j\alpha}^* = V^{\dagger}_{\alpha
  j}$ which transform the many-body bases for the local fragment A and
its environment B separately. If $N_B$ is larger than $N_A$, this
Schmidt decomposition shows that we only need to retain at most $N_A$
many-body states for the environment B, in order to express our
desired $\ket{\Psi}$.

The $N_A$ many-body Schmidt states $\ket{\widetilde{B}_\alpha}$ define
an exact DMET bath for the fragment A. If $\ket{\Psi}$ is the
ground-state of a Hamiltonian $H$ in the full system, then it is {\it
  also} the ground-state of the embedding Hamiltonian
\begin{align}
H_{emb} = P H P \label{projected_hamiltonian}
\end{align}
where $P = \sum_{\alpha \beta} \ket{\widetilde{A}_\alpha
  \widetilde{B}_\beta} \bra{\widetilde{A}_\alpha
  \widetilde{B}_\beta}$. This is the heart of the DMET construction:
the solution of a small embedded problem, consisting of a fragment
plus its bath, is the same as the solution of the full system.

In practice, DMET approximations must enter however, because the bath
construction itself requires the solution state $\ket{\Psi}$. DMET is
thus formulated in a boot-strap manner, where an {\it approximate}
low-level $\ket{\Phi}$ for the full system is first used to derive the
DMET bath, and then improved self-consistently from the high-level
solution of the small embedded problem, which yields a high-level
$\ket{\Psi}$. Different DMET approximations in the literature use
different states $\ket{\Phi}$ and impose different forms of
self-consistency between $\ket{\Psi}$ and $\ket{\Phi}$.

\begin{figure}
 \includegraphics[width=0.2\textwidth]{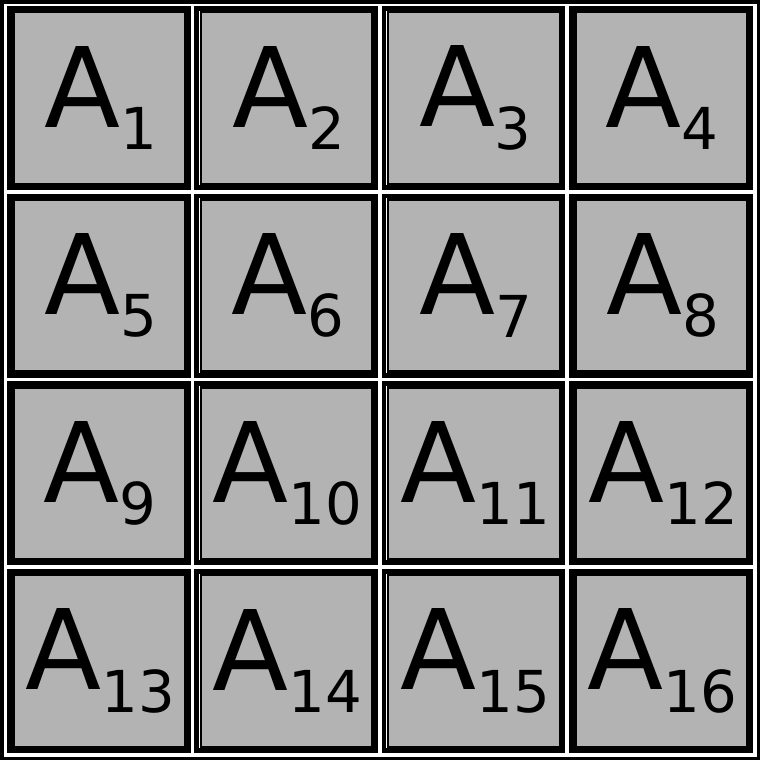}
 \caption{\label{many_fragments_fig} Division of the universe into
   local fragments.}
\end{figure}

In a general DMET calculation, the total system can be divided into
multiple local fragments, see e.g. Fig. \ref{many_fragments_fig}.  In
this case, each local fragment A$_x$ is associated with its own
embedded problem and high-level wavefunction
$\ket{\Psi_x}$. Consistency between the different $\ket{\Psi_x}$ must
then be enforced. This is carried out via self-consistency with a {\it
  single} low-level $\ket{\Phi}$ used to describe the total system.

Various kinds of low-level wavefunctions have been explored in the
literature. These include wavefunctions with correlation, such as
configuration interaction wavefunctions in Ref. \citenum{georgePRB},
block product states for spins in Ref. \citenum{spinsystemPRB}, and
AGP wavefunctions in Ref. \citenum{troyGeminals}. These forms of
$\ket{\Phi}$ yield correlated many-body Schmidt states, whose matrix
elements must be explicitly computed in the embedding Hamiltonian.
However, although there are real benefits to using the most accurate
feasible $\ket{\Phi}$ in the bath construction, it is also convenient
to recycle the large number of existing quantum many-body solvers when
solving the embedded problem.  When a low-level wavefunction of
mean-field form is used, such as a Slater determinant, the $N_A$
many-body states for the environment B are spanned by a Fock space of
single-particle states, equal in number to the number of
single-particle states of the local fragment A.\cite{geraldJCTC} This
orbital representation of the bath then allows us to reuse existing
quantum many-body solvers with little modification.  In this work, we
will focus therefore on low-level Slater determinant wavefunctions.

\subsection{Bath \textit{orbitals} from a Slater determinant}

Consider a Slater determinant approximation $\ket{\Phi_0}$ for the
ground-state of the full system. In second quantization, it can be
written as
\begin{equation}
\ket{\Phi_0} = \prod\limits_{\mu \in \text{occ}} \hat{a}^{\dagger}_{\mu} \ket{-}. \label{HFwfn}
\end{equation}
Here, $\mu$ denotes occupied spin-orbitals and $\ket{-}$ denotes the
true vacuum. In lattice model language, the \textit{spin-orbital}
indices combine the lattice site and spin indices into one index. A
spin-orbital therefore has two possible occupations. \textit{Spatial
  orbitals} correspond to the lattice sites which have four possible
occupations. In what follows, we always assume orthonormal
spin-orbitals for the local fragment and its environment. They will be
denoted by $k,l,m,n$ and there are $L$ of them. The occupied orbitals
are of course always orthonormal. They will be denoted by $\mu,\nu$
and there are $N_{\text{occ}}$ of them. The orthonormal local fragment
and bath orbitals will be denoted by $p,q,r,s$. There are $L_A$
orbitals in the local fragment A.

The occupied orbitals can be written in terms of the local fragment
and environment orbitals:
\begin{equation}
\hat{a}^{\dagger}_{\mu} = \sum\limits_{k \in AB} \hat{a}_k^{\dagger} C_{k\mu},
\end{equation}
The physical wavefunction represented by Eq. \eqref{HFwfn} does not
change when the occupied orbitals are rotated amongst each
other.\cite{helgaker,PhysRevB.88.075122} Ref. \citenum{geraldJCTC}
discusses how this freedom can be used to split the occupied orbital
space into two parts: orbitals with and without overlap on the local
fragment. This construction can be understood by means of a singular
value decomposition. Consider the occupied orbital coefficient block
with indices on the local fragment: $k \in A$. The singular value
decomposition of the $L_A \times N_{\text{occ}}$ coefficient block
$C_{k\mu}$ yields an occupied orbital rotation matrix $V_{\mu p}$:
\begin{equation}
C_{k\mu}(k \in A) = \sum\limits_{p}^{L_A} U_{kp} \lambda_{p} V^{\dagger}_{p\mu},
\end{equation}
which can be made square by adding $N_{\text{occ}}-L_A$ extra columns:
$W = \left[ V \widetilde{V} \right]$. The occupied orbital space can
now be rotated with the $N_{\text{occ}} \times N_{\text{occ}}$ matrix
$W$:
\begin{equation}
\hat{a}^{\dagger}_{p} = \sum\limits_{\mu}^{N_{\text{occ}}} \hat{a}^{\dagger}_{\mu} W_{\mu p} = \sum\limits_{\mu}^{N_{\text{occ}}} \sum\limits_{k \in AB} \hat{a}_k^{\dagger} C_{k\mu} W_{\mu p} = \sum\limits_{k \in AB} \hat{a}_k^{\dagger} \widetilde{C}_{kp}.\label{bathorbsconstructattempt1}
\end{equation}
Of the rotated occupied orbitals, only $L_A$ have nonzero overlap with
the local fragment:
\begin{equation}
\widetilde{C}_{kp}(k \in A) = \sum\limits_{\mu}^{N_{\text{occ}}} \sum\limits_{q}^{L_A} U_{kq} \lambda_{q}  V^{\dagger}_{q\mu} W_{\mu p} =
\begin{cases} 
   U_{kp} \lambda_{p} & \text{if } p \leq L_A \\
   0       & \text{otherwise}
  \end{cases}.
\end{equation}
This construction assumes that $L_A \leq N_{\text{occ}}$. This
assumption can fail when we use large basis sets in quantum
chemistry. We return to this issue in
Sec. \ref{practicalbathorbitalconstructionsection}.

The Schmidt eigenstates $\left\{ \ket{\widetilde{B}_{\alpha}}
\right\}$ in Eq. \eqref{schmidtdecomp} can be found by diagonalizing
the reduced density matrix of the environment B:
\begin{equation}
\hat{\rho}_B = \text{Tr}_A \ket{\Psi_0} \bra{\Psi_0} = \sum\limits_{i}^{N_A} \braket{A_i \mid \Psi_0} \braket{\Psi_0 \mid A_i} = \sum\limits_{\alpha}^{\min(N_A,N_B)} \lambda_{\alpha}^2 \ket{\widetilde{B}_{\alpha}} \bra{\widetilde{B}_{\alpha}}. \label{manybodybasisstateconstruction}
\end{equation}

Consider $\left\{ \braket{A_i \mid \Phi_0} \right\}$, the overlap of
the Slater determinant with the many-body basis states of the local
fragment A. The Slater determinant can be factorized into two parts:
one part which contains the orbitals with overlap on the local
fragment and a second part which contains the orbitals without overlap
on the local fragment:
\begin{equation}
\ket{\Phi_0} = \left( \prod\limits_{p \leq L_A} \hat{a}^{\dagger}_{p} \right) \left( \prod\limits_{L_A < p \leq N_{\text{occ}}} \hat{a}^{\dagger}_{p} \right) \ket{-}. \label{HFwfn2}
\end{equation}
The $N_A$ states $\left\{ \ket{\widetilde{B}_{\alpha}} \right\}$ are
therefore spanned by the direct product space of (a) the
$N_{\text{occ}} - L_A$ occupied orbitals without overlap on the local
fragment and (b) the Fock space consisting of the $L_A$ entangled
orbitals with overlap on the local fragment, after they have been
projected onto the environment. The construction in Eq. (5) of
Ref. \citenum{geraldJCTC} is based on the overlap of the occupied
orbitals with the local fragment:
\begin{equation}
S_{\mu\nu} = \sum\limits_{k \in A} C^{\dagger}_{\mu k}C_{k\nu} = \sum\limits_{p}^{L_A} V_{\mu p} \lambda_{p}^2 V^{\dagger}_{p \nu}. \label{origOverlap}
\end{equation}
It is immediately clear from the discussion above that at most $L_A$
eigenvalues of $S_{\mu\nu}$ are nonzero. The $L_A$ corresponding
eigenvectors yield the bath orbitals ($r \leq L_A$):
\begin{equation}
\hat{a}^{\dagger}_{r} = \sum\limits_{k \in B} \hat{a}_{k}^{\dagger} \frac{ \widetilde{C}_{k r} }{ \sqrt{\sum\limits_{l \in B} \mid \widetilde{C}_{l r} \mid^2 } } = \sum\limits_{k \in B} \sum\limits_{\mu}^{N_{\text{occ}}} \hat{a}_{k}^{\dagger} \frac{ C_{k \mu} V_{\mu r} }{ \sqrt{ 1 - \lambda_r^2 } }. \label{bathorbitalconstruction}
\end{equation}
The bath orbitals in Eq. \eqref{bathorbitalconstruction} are exactly
those from Eq. \eqref{bathorbsconstructattempt1}, after the latter
have been projected onto the environment.

\begin{figure}
 \includegraphics[width=0.35\textwidth]{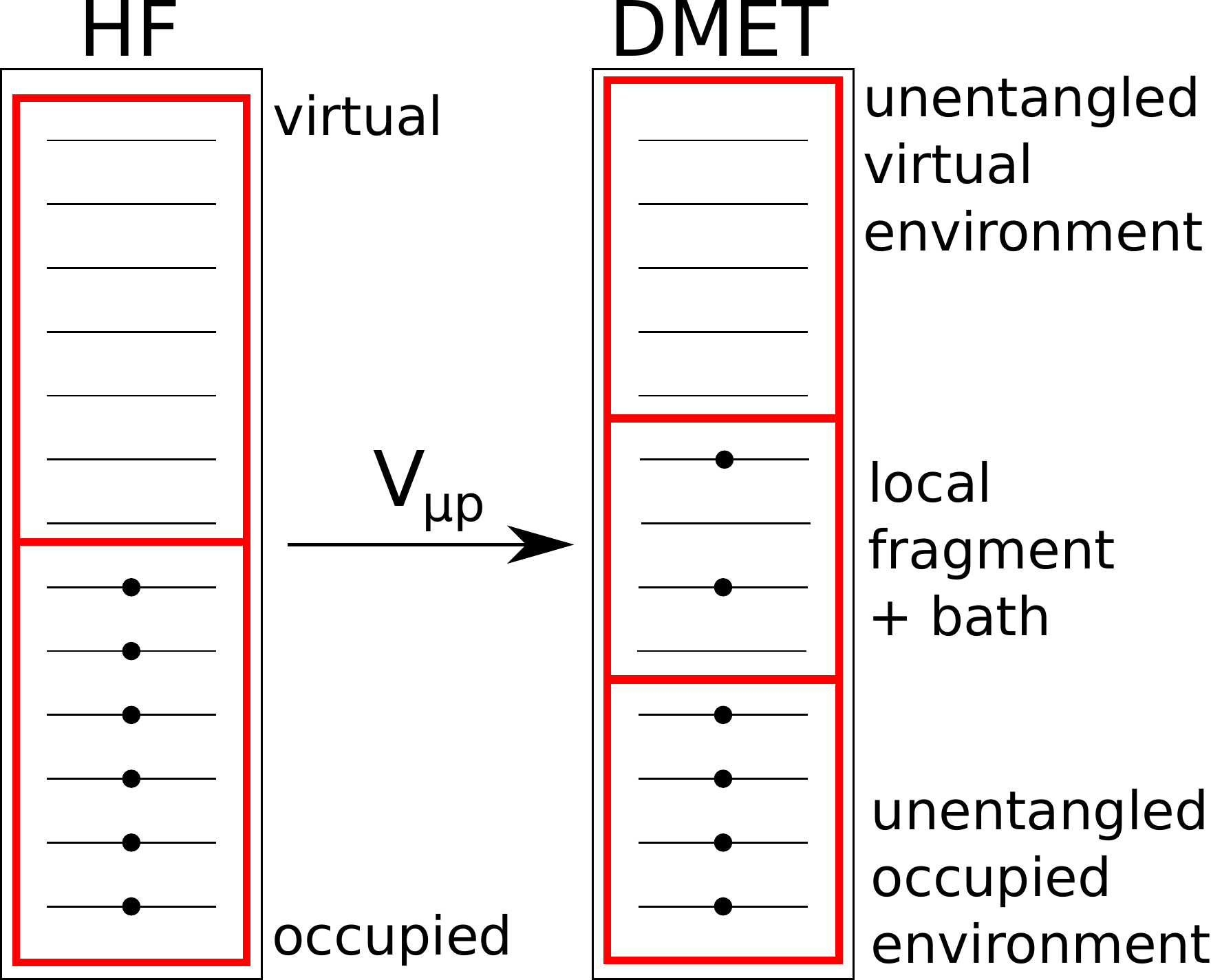}
 \caption{\label{activeSpaceHamFig} The bath orbital transformation
   generates an active space. Note that the depicted ordering of the
   orbitals is arbitrary.}
\end{figure}

From the above, we see that the DMET construction yields 4 kinds of
orbitals: local fragment orbitals, bath orbitals, unentangled occupied
environment orbitals, and unentangled unoccupied environment
orbitals. The bath orbitals and local fragment orbitals will in
general be partially occupied in the DMET high-level wavefunction
$\ket{\Psi}$, thus the bath plus local fragment space is a quantum
chemistry {\it active space}. In active space language, the
unentangled occupied environment orbitals are external core orbitals,
and the unentangled unoccupied environment orbitals are external
virtual orbitals.  Fig. \ref{activeSpaceHamFig} illustrates the
relationship of the original basis to the active space representation
generated by the transformation to bath orbitals.  Because there are
$N_{\text{occ}} - L_A$ core orbitals, the active space of $2L_A$ local
fragment and bath orbitals will contain precisely $L_A$ electrons.
The $N_{\text{occ}} - L_A$ core orbitals can contribute direct and
exchange terms to the embedded Hamiltonian, as discussed in
Sec. \ref{active_space_ahm_major_Section}.

\subsection{Bath orbital construction in practice}\label{practicalbathorbitalconstructionsection}

We now present a second and completely equivalent construction of the
bath orbitals. In this formulation, only the mean-field density matrix
in the local fragment and environment orbital basis is required:
\begin{equation}
D_{kl} = \braket{\Phi_0 \mid \hat{a}_l^{\dagger} \hat{a}_k \mid \Phi_0} = \sum\limits_{\mu}^{N_{\text{occ}}} C_{k \mu} C^{\dagger}_{\mu l} = \sum\limits_{p}^{N_{\text{occ}}} \widetilde{C}_{k p} \widetilde{C}^{\dagger}_{p l}.
\end{equation}
The eigenvalues of this idempotent density matrix are all either 0 or
1. Consider the $(L - L_A) \times (L - L_A)$ subblock where $k$ and
$l$ belong to the environment B only. We have hence removed the $L_A$
rows and columns corresponding to the local fragment. Due to
MacDonald's theorem,\cite{PhysRev.43.830} at most $L_A$ eigenvalues of
the $(L - L_A) \times (L - L_A)$ subblock will lie in between 0 and
1. The corresponding eigenvectors are the orthonormal bath orbitals
from Eq. \eqref{bathorbitalconstruction}. The $N_{\text{occ}} - L_A$
eigenvectors with eigenvalue 1 are the unentangled occupied
environment orbitals, which give direct and exchange contributions to
the active space Hamiltonian, see
Sec. \ref{active_space_ahm_major_Section}.

The overlap matrix $S_{\mu\nu}$ in Eq. \eqref{origOverlap} is a
projector of the occupied orbitals onto the local
fragment. Analogously, $D_{kl}(kl \in B)$ is a projector of the
environment orbitals onto the occupied orbitals. Any eigenvectors with
partial weight signal occupied orbitals with support on both the local
fragment and the environment, i.e. the entangled occupied orbitals.

In practical calculations in quantum chemistry, the selection of bath
orbitals is intimately tied to the localization procedure used to
determine how orbitals define fragments. One possibility is to
localize orbitals using some standard procedure (L\"owdin
orthogonalization, Boys localization, etc.) and defining the fragments
accordingly. It is important to note, however, that the localization
must mix particle and hole states so that at least some of the
fragment orbitals become entangled. If this strategy is followed, some
of the $L_A$ fractional eigenvalues of $D_{kl}(kl \in B)$ can lie
arbitrarily close to 0 or 1 (or to 0 or 2 when a spin-summed
restricted Slater determinant is used as the low-level
wavefunction). It can happen for very large basis sets
$(N_{\text{occ}} < L_A)$, or when the occupied core orbitals of
neighbouring atoms are in practice unentangled. This makes it
difficult to distinguish between true bath orbitals and unentangled
environment orbitals.  In such cases, one approach is to use an
eigenvalue cutoff (e.g. $10^{-13}$) to discard the corresponding
eigenvectors from the bath orbital space.  However, in chemical
applications this truncation can lead to problems, for example, if
different sets of bath orbitals enter at different points on a
potential energy surface. A practical fix is then to keep only one
bath orbital per broken chemical bond, as was first presented in
Ref. \citenum{qimingJCTC}.

In this work, we have considered a more elaborate localization
strategy using the ideas expressed in Ref.~\citenum{iao_gerald}. In a
typical calculation, we determine fragment core orbitals by projecting
the occupied MO set into core-like AOs. In the valence space, we use
the intrinsic atomic orbital (IAO) construction described in
Ref.~\citenum{iao_gerald}. This yields a set of localized, atomic-like
orbitals that exactly span the occupied MO space. Localized,
atomic-like virtual orbitals are then determined by a projection into
a set of corresponding atomic-like orbitals (appropriately
orthogonalizing with respect to the previous sets). If this strategy
is followed, entangled orbitals in the fragment are restricted to the
valence IAO set, while core (and virtual) orbitals keep this character
within the fragment. We find this strategy closer to the spirit of
DMET and can avoid some of the arbitrariness in choosing an eigenvalue
cutoff to determine entangled orbitals. It can also provide more
consistent results as the atomic basis set is increased towards
completeness.

Due to the possibility of truncation, we will henceforth denote the
number of bath orbitals by $L_B$, where $L_B \leq L_A$.  Once the bath
orbitals are determined by diagonalizing $D_{kl}(kl \in B)$, all other
environment orbitals are restricted to be fully occupied or
empty. Thus with truncation, the deficit in electron number between
the fully occupied environment orbitals and $N_{\text{occ}}$ is the
number of electrons $N_{\text{act}}$ in the active space.

\section{The DMET Hamiltonians and self-consistency} \label{active_space_ahm_major_Section}

We now introduce the low-level and high-level DMET Hamiltonians which
are connected by the DMET correlation potential and self-consistency.
In lattice applications of DMET, the low-level Hamiltonian is termed
the lattice Hamiltonian, and the high-level embedding Hamiltonian is
termed the impurity Hamiltonian. As the quantities are all related, we
discuss in general terms their role in the theory, before we give
their precise definition.

We first start with the Hamiltonian for the total system. For a
general chemical problem, it takes the form:
\begin{equation}
\hat{H} = E_{\text{nuc}} + \sum\limits_{kl}^L t_{kl} \hat{a}_k^{\dagger} \hat{a}_l + \frac{1}{2} \sum\limits_{klmn}^L (kl|mn) \hat{a}_k^{\dagger} \hat{a}_m^{\dagger}  \hat{a}_n \hat{a}_l, \label{orig_ham_eq}
\end{equation}
where $t_{kl}$ and $(kl|mn)$ are diagonal in the spin indices of
spin-orbitals $k$ and $l$, and $(kl|mn)$ is diagonal in the spin
indices of $m$ and $n$ as well. Throughout this work, fourfold
permutation symmetry $(kl|mn) = (mn|kl) = (lk|nm)$ is assumed.

$\hat{H}$ yields $\ket{\Psi_0}$ as its exact ground state, as appears
in Eq. \eqref{schmidtdecomp}. However, in DMET, the determination of
$\ket{\Psi_0}$ is replaced by the determination of a low-level
wavefunction $\ket{\Phi}$ and a set of high-level wavefunctions
$\ket{\Psi_x}$.  $\ket{\Phi}$ is the ground-state of the DMET
low-level Hamiltonian $\hat{h}'$ (vide infra), and $\ket{\Psi_x}$ is
the ground-state of the DMET high-level (embedding) Hamiltonian
$\hat{H}_{emb}^{x}$ for fragment A$_x$. Both are derived from
$\hat{H}$, and are connected by the correlation potential
$\hat{C}$. The correlation potential $\hat{C}$ is adjusted to match
observables in $\ket{\Phi}$ and $\ket{\Psi}$ through a
self-consistency cycle. As we shall see, the form of $\hat{C}$ depends
on the observables we choose to match, and the type of $\ket{\Phi}$ we
are using.

In the formulation we focus on here, the low-level $\ket{\Phi}$ is a
Slater determinant. Thus $\hat{h}'$ is a one-particle Hamiltonian of
the form
\begin{equation}
\hat{h}' = \hat{h} + \sum_x \hat{C}_x \label{mean-field-operator-eq}
\end{equation}
where $\hat{C}$ is a sum of one-particle operators  acting on each of the blocks of fragment orbitals,
\begin{equation}
\hat{C}_x = \sum\limits_{kl \in A_x} u^x_{kl} \hat{a}^{\dagger}_k \hat{a}_l, \label{correlationpotent_eq}
\end{equation}
and the $u^x_{kl}$ matrix elements are adjusted to match
single-particle density matrices $\langle a^\dag_p a_q \rangle$
between the high-level wavefunction $\ket{\Psi_x}$ and the global
low-level wavefunction $\ket{\Phi}$. We delay the precise description
of the matching until
Sec.~\ref{correlationpotentialoptimization_subsection}, but note that
unless there is translational invariance or other symmetries to relate
the fragments, $\hat{C}_x$ will be different for each fragment.
$\hat{h}$ is a single-particle Hamiltonian, which may be held fixed
along the DMET optimization. The simplest choice for $\hat{h}$ is the
one-particle part of the total Hamiltonian $\hat{H}$, and in this case
one relies on the correlation potential $\hat{C}$ to produce the
mean-field Fock-like Coulomb and exchange contributions on the
fragments, as the correlation potential is adjusted by the
self-consistency.  Alternatively, one can choose the initial $\hat{h}$
to be the Fock operator $\hat{F}$ derived from $\hat{H}$.  In this
case, however, the Coulomb and exchange potentials of $\hat{F}$ that
act in each fragment A$_x$ will be redundant with $\hat{C}_x$ during
the self-consistency, although the components that act outside of the
fragments are not.  If $\hat{H}$ only has Coulomb terms which act in
each fragment separately (as in the Hubbard model) choosing $\hat{h}$
to be the Fock operator or the hopping Hamiltonian $\sum\limits_{kl}^L
t_{kl} \hat{a}_k^{\dagger} \hat{a}_l$ is {\it exactly
  equivalent}. Thus in applications of DMET to the Hubbard model, the
simpler hopping Hamiltonian $\hat{h}$ has been commonly used.

We now discuss the high-level embedding Hamiltonians. There are two
choices: an interacting bath high-level Hamiltonian, and a
non-interacting bath high-level Hamiltonian.

\subsection{Interacting bath formulation}
The high-level embedding Hamiltonian $\hat{H}_{emb}^x$ is an
interacting Hamiltonian for the active space of local fragment A$_x$.
The conceptually simplest construction of $\hat{H}_{emb}^x$ is to
project the total Hamiltonian $\hat{H}$ into the active space
representation of fragment A$_x$, as in
Eq. \eqref{projected_hamiltonian}. We can do this by writing the
one-particle part of $\hat{H}_{emb}^x$ as
\begin{align}
\hat{h}^x = \sum_{kl} \left[ t_{kl} + \sum\limits_{mn}^{L} \left[ (kl|mn) - (kn|ml) \right] D_{mn}^{\text{env,x}} \right] \hat{a}_k^\dag \hat{a}_l = \sum_{kl} \widetilde{h}^x_{kl} \hat{a}_k^\dag \hat{a}_l, \label{lskhhfhvhvhhehdhdhdh}
\end{align}
(note the inclusion of the Coulomb and exchange terms from the
unentangled occupied environment), and then transforming the
one-particle part to the active representation of fragment A$_x$, and
adding the active space two-electron integrals, yielding
\begin{align}
\hat{H}_{emb}^{x} = P \hat{H} P = \sum\limits_{rs}^{L_{A_x} + L_{B_x}} \widetilde{h}^x_{rs} \hat{a}_r^{\dagger} \hat{a}_s 
 +
\sum\limits_{pqrs}^{L_{A_x} + L_{B_x}} (pq|rs) \hat{a}_p^{\dagger} \hat{a}_r^{\dagger}  \hat{a}_s \hat{a}_q,
\end{align}
where $P$ denotes the transformation and projection into the active
space of fragment A$_x$.  Note that the correlation potential does not
actually appear in $\hat{H}_{emb}^x$, as this would double-count the
effects of the interactions already included in the active space.  The
correlation potential appears only indirectly through its effect on
the form of the bath and core orbitals. However, to ensure that the
total number of electrons in all local fragments adds up to
$N_{\text{occ}}$, it becomes necessary to introduce a global chemical
potential for the local fragment orbitals, thus giving
\begin{align}
\hat{H}^{emb}_{x} \longleftarrow \hat{H}^{emb}_{x} - \mu_{\text{glob}} \sum\limits_{r \in A_x}  \hat{a}_r^{\dagger} \hat{a}_r. \label{ihamiltonian}
\end{align}
Note that $\mu_{\text{glob}}$ does not depend on orbital ($r$) or
fragment ($x$) indices.

\subsection{Non-interacting bath formulation}

A simpler construction, motivated by the impurity formulation of
dynamical mean-field theory, can also be used.  Here Coulomb
interactions are only included on the fragment orbitals while the
correlation potential $\hat{C}$ is used to mimic the Coulomb
interactions elsewhere. This is known as the {\it non-interacting
  bath} formulation of DMET.  Here we first define the single-particle
part of the high-level Hamiltonian as
\begin{align}
\hat{h}^x = \hat{h} + \sum_{x \neq A} \hat{C}_x \label{lskhhfhvhvhhehdhdhdh2}
\end{align}
where we observe that the correlation potential appears on all sites
{\it outside} of the fragment.  We then transform this to the fragment
plus bath representation of fragment A$_x$ and include the
two-particle interactions {\it only on the fragment orbitals}, giving
(including a chemical potential)
\begin{align}
\hat{H}_{emb}^x = \sum\limits_{rs}^{L_{A_x} + L_{B_x}} \widetilde{h}^x_{rs} \hat{a}_r^{\dagger} \hat{a}_s - \mu_{\text{glob}} \sum\limits_{r \in A_x}  \hat{a}_r^{\dagger} \hat{a}_r + \sum\limits_{klmn}^{L_{A_x}} (kl|mn) \hat{a}_k^{\dagger} \hat{a}_m^{\dagger}  \hat{a}_n \hat{a}_l. \label{nihamiltonian}
\end{align}
Because the correlation potential appears directly in the high-level
Hamiltonian through its contribution to
Eq. \eqref{lskhhfhvhvhhehdhdhdh2}, the correlation potential can
itself be used control the total particle number in all the local
fragments. Thus if matching the particle number between $\ket{\Phi}$
and the union of all $\ket{\Psi_x}$ is achieved perfectly in the DMET
self-consistency cycle, the chemical potential $\mu_{\text{glob}}$
appearing in Eq.~\eqref{nihamiltonian} is redundant and can be
omitted. Alternatively, $\hat{C}$ (i.e. $u^x_{kl}$) can be constrained
to be traceless, and then $\mu_{\text{glob}}$ takes on the meaning of
the diagonal part of $\hat{C}$. We typically use the latter strategy.

While the non-interacting bath formulation only includes two-particle
interactions on the fragment orbitals, it nonetheless converges to the
exact result as the fragment size increases. Thus either the
non-interacting bath or interacting bath formulation can be used and
may be convenient for different purposes.  For example, the first
studies of the Hubbard model with DMET used the non-interacting bath
formulation, because many quantum Monte Carlo methods used in this
problem have difficulty with the non-local two-electron interactions
that appear in the interacting bath formulation of DMET.
However, for quantum chemical solvers such as configuration
interaction or coupled cluster theory, the only benefit to omitting
the bath two-particle interactions is to reduce the number of
two-electron integrals to compute. This is not a large advantage in
practice, when weighed against neglecting the bath correlations.  For
this reason, we will focus on the interacting bath formulation in the
calculations in this work.

\section{DMET expectation values} \label{DMETenergy_major_seciton}

The ground-state of each fragment DMET high-level Hamiltonian yields a
high-level wavefunction $\ket{\Psi_x}$. These high-level wavefunctions
are used to assemble the DMET expectation values of interest. Note
that if the fragments are non-overlapping as is typical in DMET, each
fragment wavefunction defines the expectation values for operators
that act {\it locally} on each fragment. For example, from each local
fragment's $\ket{\Psi_x}$, we obtain the one-particle and two-particle
density matrices (1- and 2-RDM) on the fragments
\begin{eqnarray}
D_{sr}^x & = & \braket{ \hat{a}^{\dagger}_r \hat{a}_s }, \\
P^x_{qp|sr} & = & \braket{ \hat{a}^{\dagger}_p \hat{a}^{\dagger}_r \hat{a}_s \hat{a}_q },
\end{eqnarray}
with $pqrs \in {A_x}$. We refer to this as a ``democratic'' evaluation
of the local expectation values.

For non-local operators that act on multiple fragments, different
fragments' high-level wavefunctions will in general yield different
values for the non-local expectation values. Clearly it is desirable
to combine the expectation values from different fragments in an
optimal way. The solution used in the original DMET was to partition
the expectation value of a Hermitian sum of non-local operators, such
as $\hat{a}_i^\dag \hat{a}_j + \hat{a}_j^\dag \hat{a}_i$, in a
similarly democratic fashion as
\begin{align}
\langle \hat{a}_i^\dag \hat{a}_j +  \hat{a}_j^\dag \hat{a}_i\rangle  = 
\langle \Psi_{x(i)} | \hat{a}_i^\dag \hat{a}_j | \Psi_{x(i)}\rangle + 
\langle \Psi_{x(j)} | \hat{a}_j^\dag \hat{a}_i | \Psi_{x(j)}\rangle 
\end{align}
where $x(i)$ denotes the fragment containing orbital $i$, i.e.  the
first index of the operator determined the fragment wavefunction to
use.  Following this rule, the total energy corresponding to the
Hamiltonian $\hat{H}$ in Eq.~(\ref{orig_ham_eq}) is evaluated as
\begin{align}
E_{\text{tot}} & =  E_{\text{nuc}} + \sum\limits_x E_{x}, \\
E_{x} & =  \sum\limits_{k \in A_x} \left( \sum\limits_{l}^L t_{kl} D_{lk}^{\text{tot}} + \frac{1}{2} \sum\limits_{lmn}^L (kl|mn) P_{lk|nm}^{\text{tot}} \right). \label{energyfragment}
\end{align}

However, there are some cases where this ``democratic'' partitioning
of non-local expectation values is sub-optimal. This can be observed
when a single fragment (labeled by A) is treated with a high-level
method while other fragments are treated at a lower level of
theory. In this case, the non-local expectation values associated with
the high-level wavefunction of fragment A are more accurate than the
expectation values associated with the low-level wavefunction of other
fragments. Then, it is more accurate to define
\begin{align}
\langle \hat{a}_i^\dag \hat{a}_j + \hat{a}_j^{\dag} a_i \rangle = \langle \Psi_A | \hat{a}^\dag_i  \hat{a}_j | \Psi_A\rangle + \langle \Psi_A | \hat{a}^\dag_j  \hat{a}_i | \Psi_A\rangle.
\end{align}
In the extreme case where a single fragment is treated at a high level
of theory while other fragments are treated at the same level of
theory as that used to obtain the Slater determinant $|\Phi \rangle$,
then it is more accurate to define {\it all} expectation values using
the high-level wavefunction for fragment A. In this case, the energy
expression becomes
\begin{align}
E_{\text{tot}}  = E_{\text{nuc}} +  E_{A}.
\label{active_space_energy}
\end{align}
We will see an example of this in the applications section.

It is important to note that not only the fragment and bath orbitals,
but also the {\it core} (unentangled occupied environment) orbitals in
$\ket{\Psi_x}$ contribute to non-local expectation values. For
example, in Eq.~(\ref{energyfragment}) the density matrices are {\it
  total} density matrices including the core contributions. Not
including the core contributions leads to inaccurate values for
non-local expectation values. This can be seen in
Ref.~\citenum{bulikPRB}, where the non-local correlation functions did
not use the core contributions. For the interacting bath formulation
with democratic partitioning, the fragment energies
\eqref{energyfragment} become
\begin{equation}
E_{x} \approx \sum\limits_{p \in A_x} \left( \sum\limits_{q}^{L_{A_x}+L_{B_x}} \frac{t_{pq} + \widetilde{h}_{pq}^x}{2} D_{qp}^{x} + \frac{1}{2} \sum\limits_{qrs}^{L_{A_x}+L_{B_x}} (pq|rs) P_{qp|sr}^{x} \right), \label{poiuytrewqazxcvbnmkl}
\end{equation}
with $\widetilde{h}_{pq}^x$ the rotated one-electron integrals from
Eq. \eqref{lskhhfhvhvhhehdhdhdh}. The one-electron integrals in
Eq. \eqref{poiuytrewqazxcvbnmkl} avoid the double counting of Coulomb
and exchange contributions of the \textit{core} (unentangled occupied
environment) orbitals. The factor $\frac{1}{2}$ is similar to the
difference between the Fock operator and energy expressions in HF
theory.

\section{Optimization of the low-level Hamiltonian and correlation potential} \label{correlationpotentialoptimization_subsection}

The final component in the DMET algorithm is to determine the
correlation potential $\hat{C}$. As introduced above, in the
interacting bath formulation the correlation potential appears in the
low-level Hamiltonian $\hat{h}'$, while in the non-interacting bath
formulation, it appears in both the low-level Hamiltonian $\hat{h}'$
and high-level Hamiltonian $\hat{H}_{emb}^{x}$.

Optimizing the correlation potential requires choosing observables to
match between the low-level and high-level wavefunctions, which
defines an associated cost function.  Different cost functions lead to
different DMET functional constructions, with different properties.
For example, matching the density matrices of the fragments makes the
DMET observables a functional of the self-consistently converged
density matrices: DMET is then a local density matrix functional
theory.  Matching only the diagonal elements of the density matrices
in DMET similarly provides a lattice density functional interpretation
of DMET.

Some of the cost functions and correlation potential forms that have
been used in DMET calculations include: matching the full density
matrix of the fragment plus bath orbitals (but using correlation
potentials defined in the usual way on the fragments only, $u^x_{kl}$
for $kl \in A_x$ in Eq.~(\ref{correlationpotent_eq})), which we term
fragment (impurity) plus bath fitting,\cite{geraldPRL, geraldJCTC}
matching the density matrix of the fragments only using correlation
potentials on the fragments, which we term fragment (impurity) only
fitting,\cite{geraldPRL, geraldJCTC} matching the diagonals of the
fragment density matrices, using diagonal correlation potentials on
the fragments ($u^x_{kl} = u^x_{kk}\delta_{kl}$),\cite{bulikPRB} and
using no correlation potential ($u^x_{kl}=0$) and only matching the
total number of electrons with a global chemical potential
$\mu_{\text{glob}}$,\cite{bulikJCP} which we refer to as single-shot
embedding.  Written explicitly, these cost functions are respectively,
for the fragment plus bath density matrices:\cite{geraldPRL}
\begin{equation}
\text{CF}_{\text{full}}(u) = \sum\limits_x \sum\limits_{rs}^{L_{A_x} + L_{B_x}} \left( D^x_{rs} - D^{\text{mf}}_{rs}(u) \right)^2, \label{costfunction1}
\end{equation}
the  fragment only density matrices:\cite{geraldJCTC, troyGeminals}
\begin{equation}
\text{CF}_{\text{frag}}(u) = \sum\limits_x \sum\limits_{rs \in A_x} \left( D^x_{rs} - D^{\text{mf}}_{rs}(u) \right)^2, \label{costfunction2}
\end{equation}
the  fragment only densities:\cite{bulikPRB}
\begin{equation}
\text{CF}_{\text{dens}}(u) = \sum\limits_x \sum\limits_{r \in A_x} \left( D^x_{rr} - D^{\text{mf}}_{rr}(u) \right)^2, \label{costfunction3}
\end{equation}
and for the total electron number:\cite{bulikJCP}
\begin{equation}
\text{CF}_{\text{elec}}(\mu_{\text{glob}}) = \left( \sum\limits_x \sum\limits_{r \in A_x} D^x_{rr}(\mu_{\text{glob}}) - N_{\text{occ}} \right)^2. \label{costfunction4}
\end{equation}
The latter corresponds to a global chemical potential optimization.
As discussed extensively in Ref. \citenum{troyGeminals}, trying to
mimic (parts of) a high-level correlated density matrix by (parts of)
a mean-field density matrix is not always possible because the latter
is idempotent while the former does not have to be. This is analogous
to certain densities not being non-interacting $v$-representable in
Kohn-Sham density functional theory. In such cases, the cost functions
will not minimize to zero, and this is in fact always the case for the
first cost function Eq.~(\ref{costfunction1}).  In the calculations in
this work, we focus primarily on the local fragment (``impurity
only'') density matrix matching, as in the original quantum chemistry
DMET~\cite{geraldJCTC}.

The cost function optimization algorithm in
Refs.~\citenum{geraldPRL},~\citenum{geraldJCTC} optimized the
correlation potential $u^x_{kl}$ for each local fragment A$_x$
independently.  The disadvantage is that it is prone to limit cycles
and slow convergence due to overshooting when there are multiple
fragments. Instead, we recommend to optimize the correlation potential
for all local fragments simultaneously; the stationary points of the
two procedures are the same.  We further fix the high-level density
matrix when optimizing the cost function, since then the gradient with
respect to the correlation potential can be expressed in terms of the
gradient of the low-level density matrix.  This is easily computed, as
shown in Appendix \ref{appendixGRAD}.  We then use this gradient in a
standard least-squares optimizer such as provided by \textsc{MINPACK}.
Once the new $u$ is determined the high-level density matrix is
updated.\footnote{This iterative strategy is common to all previous
  DMET works. For certain problems, it may not be an optimal numerical
  strategy similar to how fixed-point iterations in standard HF theory
  may fail in certain molecules.} The full algorithm is described in
section~\ref{sect_algo_total}.

If there exists no low-level wavefunction that exactly matches the
given (fixed) high-level density matrix fragments, the best matching
low-level density matrix $D^\text{mf}(u)$ may be undetermined during
the cost-function optimization. This is because for any non-zero value
of the cost function and fixed high-level density matrix, there is
clearly a manifold of low-level density matrices $D^\text{mf}$ (not
necessarily parametrized by $u$) which yield the same (non-zero) cost.
Indeterminacy occurs when there is a continuous intersection between
$D^\text{mf}$ and $D^\text{mf}(u)$ (i.e. the density matrices
parametrized by $u$) which is increasingly likely as the fragment size
increases and there is more freedom in $u$.  It is therefore useful to
consider an alternative formulation where this indeterminacy does not
arise.

First, consider minimizing $\langle \Phi| \hat{h} |\Phi \rangle$ under
a set of Lagrangian constraints, similar to the Kohn-Sham scheme in
density functional theory. If the local fragment density matrices are
to be matched (cf. Eq. \eqref{costfunction2}), this corresponds to the
optimization
\begin{equation}
\min_\Phi \left[ \langle \Phi | \hat{h} | \Phi \rangle + \sum\limits_x \sum\limits_{rs \in A_x} u^x_{sr} \left( D^{\text{mf}}_{rs}(\Phi) - D^x_{rs} \right)\right].  \label{ks}
\end{equation}
In Eq.~(\ref{ks}), the correlation potential
appears as the matrix of Lagrange multipliers that enforces the
constraints.  Imposing the orthonormality constraint on the orbitals
$\phi_i$ leads to a set of eigenequations satisfied at the minimum,
\begin{equation}
\sum_j \left( \hat{h} + \sum\limits_x \hat{u}^x \right)_{ij} \phi_j = \varepsilon_i \phi_i, \label{wu_eigen}
\end{equation}
where $\varepsilon_i$ are the set of Lagrange multipliers enforcing
orthonormality. This eigenvalue problem is identical to the
ground-state DMET low-level problem, where the orbitals that define
$\ket{\Phi}$ are obtained from the single-particle Hamiltonian
$\hat{h}'$. To eliminate the indeterminacy when the density matrix
constraint cannot be satisfied, we consider the dual of
Eq.~(\ref{ks}), replacing the constrained optimization by an
unconstrained maximization over the potential $u$, following Lieb
\cite{WuYang, Lieb}.  This gives the new cost function
\begin{equation}
  \text{CF}_{\text{frag}}(u) = \min_\Phi \left[ \langle \Phi | \hat{h} | \Phi \rangle + \sum\limits_x \sum\limits_{rs \in A_x} u^x_{sr} \left( D^{\text{mf}}_{rs}(\Phi) - D^x_{rs} \right) \right]. 
  \label{costfunction2_improved}
\end{equation}
In the above $\ket{\Phi}$ uses the {\em aufbau} occupations.  When an
exact match of the density matrix can be found, maximizing
Eq.~(\ref{costfunction2_improved}) is equivalent to minimizing
Eq.~(\ref{ks}), or the original cost function
Eq.~(\ref{costfunction2}).  However, the unconstrained maximization
can be performed even when no exact match exists, and then the
presence of the energy term breaks the degeneracy of imperfectly
matched solutions, removing the indeterminacy in $u$.

To maximize Eq.~(\ref{costfunction2_improved}) we use a standard BFGS
optimization (using the analytical gradients of the cost
function). For all calculations presented later using fragment density
matrix matching, it was possible to perfectly match the density
matrices, thus the cost functions Eqs. \eqref{costfunction2} or
\eqref{costfunction2_improved} gave identical results.  We also use
the mean-field Fock operator of the initial low-level wavefunction
$\ket{\Phi}$ (in practice, the restricted Hartree-Fock determinant) to
define $\hat{h}$ in Eq.~(\ref{costfunction2_improved}). This still
ties the DMET optimization problem to the original Hartree-Fock
solution. This dependence could be eliminated if the mean-field Fock
operator is also determined self-consistently. However, in the few
cases where we tried the latter we observed no noticeable difference
with respect to using the initial Fock operator.

\begin{figure*}[t!]
\begin{algorithm}[H]
  \caption{Pseudocode for the DMET algorithm}
  \label{EPSA}
   \begin{algorithmic}[1]
   \State u $\leftarrow$ 0
   \State $\mu_{\text{glob}} \leftarrow$ 0
   \Do
     \State $u_{\text{previous}} \leftarrow u$
     \State $\ket{\Phi_0(u)} \leftarrow \hat{F} + \sum\limits_x \hat{C}_x$
     \State $D_{kl}^{\text{mf}} \leftarrow \ket{\Phi_0(u)}$
     \Do
        \For{A$_x$ $\in$ system}
           \State Compute bath orbitals: $\hat{a}_r^{\dagger} \leftarrow D_{kl}^{\text{mf}}$
           \State $E_0;~h^x_{rs};~(pq|rs) \leftarrow D^{\text{env},x}_{kl} \leftarrow D_{kl}^{\text{mf}}$
           \State $D^x_{sr};~P^x_{qp|sr} \leftarrow E_0;~\widetilde{h}^x_{rs};~(pq|rs);~\mu_{\text{glob}}$
           \State $E_{x} \leftarrow D^x_{sr};~P^x_{qp|sr}$
        \EndFor
        \State $E_{\text{tot}} \leftarrow E_{\text{nuc}} + \sum\limits_x E_{x}$
        \State $N_{\text{fragments}} \leftarrow \sum\limits_x \sum\limits_{r \in A_x} D^x_{rr}$
        \State $\mu_{\text{glob}} \leftarrow \mu_{\text{glob}};~N_{\text{fragments}} - N_{\text{occ}}$
     \doWhile{$N_{\text{fragments}} \neq N_{\text{occ}}$}
     \State $u; D_{kl}^{\text{mf}} \leftarrow \min\limits_{u} \text{CF}(u)$
   \doWhile{$u \neq u_{\text{previous}}$}
   \end{algorithmic}
\end{algorithm}
\end{figure*}

\subsection{The DMET algorithm}\label{sect_algo_total}

Now that all the pieces of the DMET algorithm have been introduced,
the total DMET algorithm can be described. At the start, the system
Hamiltonian (Eq. \eqref{orig_ham_eq}) should be known, as well as the
partitioning of the system into local fragments
(Fig. \ref{many_fragments_fig}). The pseudocode for the total DMET
algorithm is given in algorithm \ref{EPSA}. On lines 4-6 the low-level
density matrix for the total system is computed for a given
correlation potential. On line 9 the DMET bath orbitals for local
fragment A$_x$ are computed according to
Sec. \ref{practicalbathorbitalconstructionsection}. On line 10 the
high-level embedding Hamiltonian for local fragment A$_x$ is
calculated according to
Sec. \ref{active_space_ahm_major_Section}. Together with a global
chemical potential which only acts on the local fragment orbitals but
not on the bath orbitals, the high-level ground state 1-RDM and 2-RDM
on line 11 are determined from the embedding Hamiltonian. The
contribution of local fragment A$_x$ to the total energy is computed
according to Eq. \eqref{energyfragment} on line 12. On line 14, these
local energy contributions are summed to yield the total DMET system
energy. On line 15, the total number of electrons in all local
fragments is obtained as a sum of local fragment traces of the
high-level 1-RDMs. When this number is different from the desired
particle number, the chemical potential needs to be adjusted. Line 16
is realized in our code by the secant root-finding method to solve for
\begin{equation}
N_{\text{fragments}}(\mu_{\text{glob}}) - N_{\text{occ}} = 0.
\end{equation}
And finally, once $\mu_{\text{glob}}$ is found, the optimization of
the correlation potential on line 18 is performed with the methods
described in Sec. \ref{correlationpotentialoptimization_subsection}.

\section{Applications} \label{section_applications}

The calculations in this work have been performed with
\textsc{qc-dmet}.\citep{github_qcdmet} The integrals in atomic and
molecular orbital spaces were obtained with
\textsc{pyscf}\citep{github_pyscf}.  As high-level methods, we have
used the coupled-cluster solver with singles and doubles (CCSD) from
\textsc{pyscf} and the FCI (DMRG) solver from
\textsc{chemps2}.\cite{chemps2_cpc} In this work, we use the CCSD {\em
  response} density matrices \cite{bartlett,ccsdrdm2} obtained using
the solution to the corresponding $\Lambda$ equations.

\begin{figure}
 \centering
 \includegraphics[trim={2cm 12.3cm 2cm 1.8cm},clip,width=1.0\textwidth]{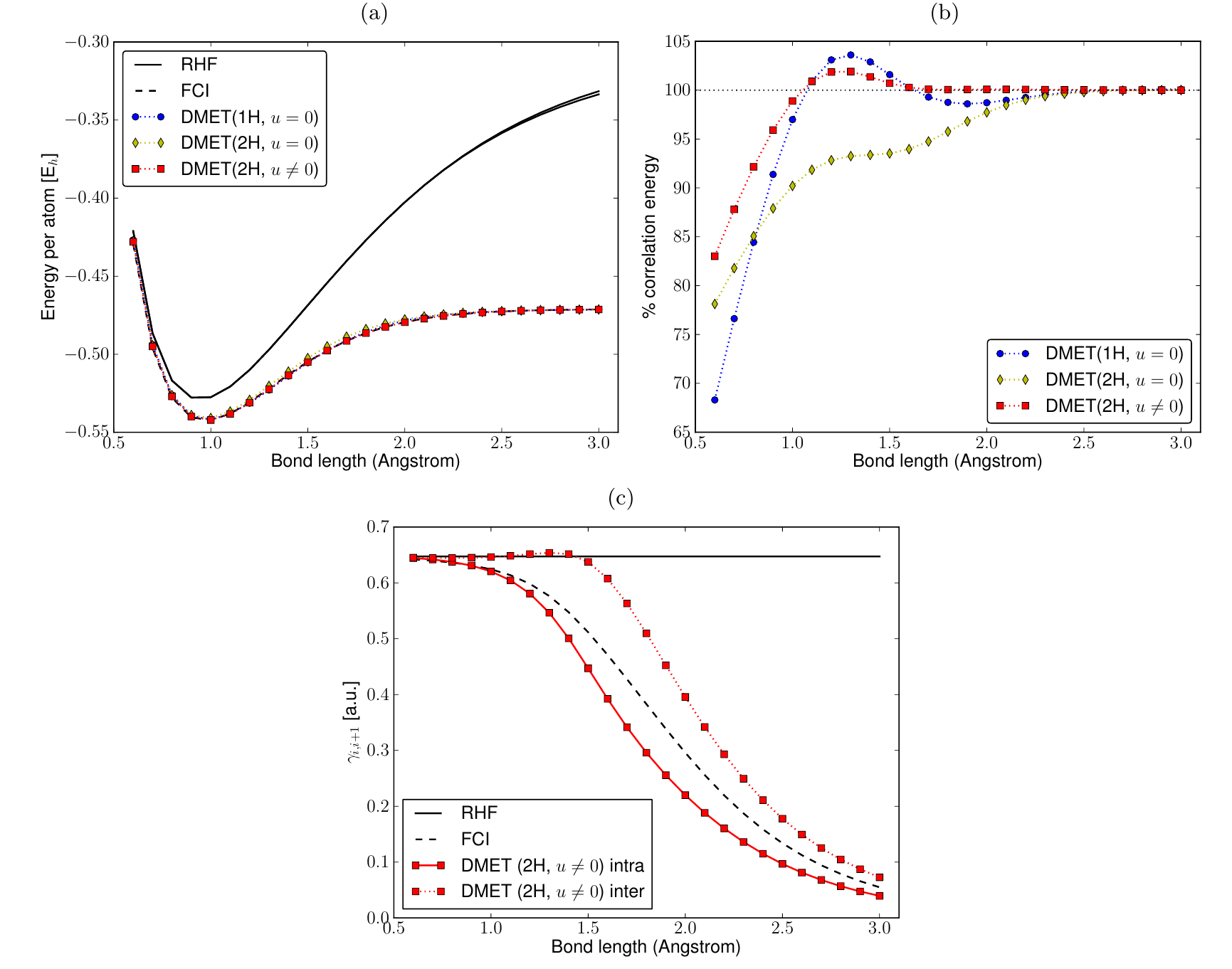}
 \caption{\label{hchain_figure} Interacting bath DMET results for the
   symmetric stretch of a hydrogen ring with 10 atoms in the STO-6G
   basis (using a L\"owdin symmetric orthogonalization). (a) Bond
   dissociation curve. Two RHF curves are displayed, corresponding to
   a fully symmetric and a dimerized solution; the corresponding
   instability occurs at $\approx 2.1$~Angstrom. (b) Fraction of the
   correlation energy captured by DMET. (c) Nearest-neighbor bond
   orders in self-consistent DMET(2H, $u \neq 0$) calculations.}
\end{figure}

\subsection{Hydrogen rings}
We present interacting bath DMET results for the symmetric stretch of
a hydrogen ring with 10 atoms in the STO-6G basis in
Fig. \ref{hchain_figure}. A L\"owdin symmetric orthogonalization was
used to define the localized orthonormal orbitals. Restricted HF was
used as the low-level method and FCI as the high-level method.
Results for fragments with one or two orbitals (one or two hydrogen
atoms) are shown. Note that due to the periodic character of the
system, only a single fragment problem needs to be solved. The fully
symmetric HF solution was used to define the DMET low-level
Hamiltonian $\hat{h}$ in self-consistent calculations; results using
the dimerized solution yield nearly indistinguishable energies. In
self-consistent calculations, we use the fragment (impurity) only
fitting (cf. Eq. \eqref{costfunction2} or
\eqref{costfunction2_improved}).

The DMET energies follow the FCI results closely along the whole
dissociation curve.\citep{geraldJCTC} More details can be observed by
plotting the fraction of correlation energy captured by DMET. The
larger deviations at smaller bond distances are due to a smaller
correlation energy, not due to larger errors in DMET.  We see that the
DMET energies are not variational, as Eq. \eqref{energyfragment} does
not correspond to the expectation value of a single wavefunction.  The
inclusion of the correlation potential (beyond the global chemical
potential) yields results significantly more accurate than without it;
in particular, nearly exact results are obtained for bond lengths
larger than 1.6~Angstrom.

We display also the nearest-neighbor bond orders ($\langle
\hat{a}^\dag_i \hat{a}_{i+1} \rangle$) in self-consistent DMET(2H)
calculations. Because of the use of 2-site fragments, two types of
bond orders can be computed in DMET: intra-fragment and
inter-fragment. In spite of their non-equivalence (which reflects the
broken full translational invariance), both are significantly improved
over RHF which remains constant (by symmetry) along the entire
dissociation curve. Note that the use of the cost function
Eq. \eqref{costfunction2} or \eqref{costfunction2_improved} guarantees
that the determinant $\ket{\Phi}$ that results from the DMET
self-consistency procedure has exactly the same intra-fragment bond
orders as the ones defined by the DMET expectation values. We see that
the DMET solution is strongly dimerized at intermediate bond
lengths. As the FCI solution must preserve translational symmetry, we
cannot detect dimerization in the single-particle density
matrix. However, the corresponding behavior might be expected in the
bond-bond correlation functions in the two-particle density
matrix. This would also indicate a tendency for the system to undergo
a Peierls transition.

\begin{figure}
 \centering
 \includegraphics[trim={2cm 19.4cm 2cm 1.8cm},clip,width=1.0\textwidth]{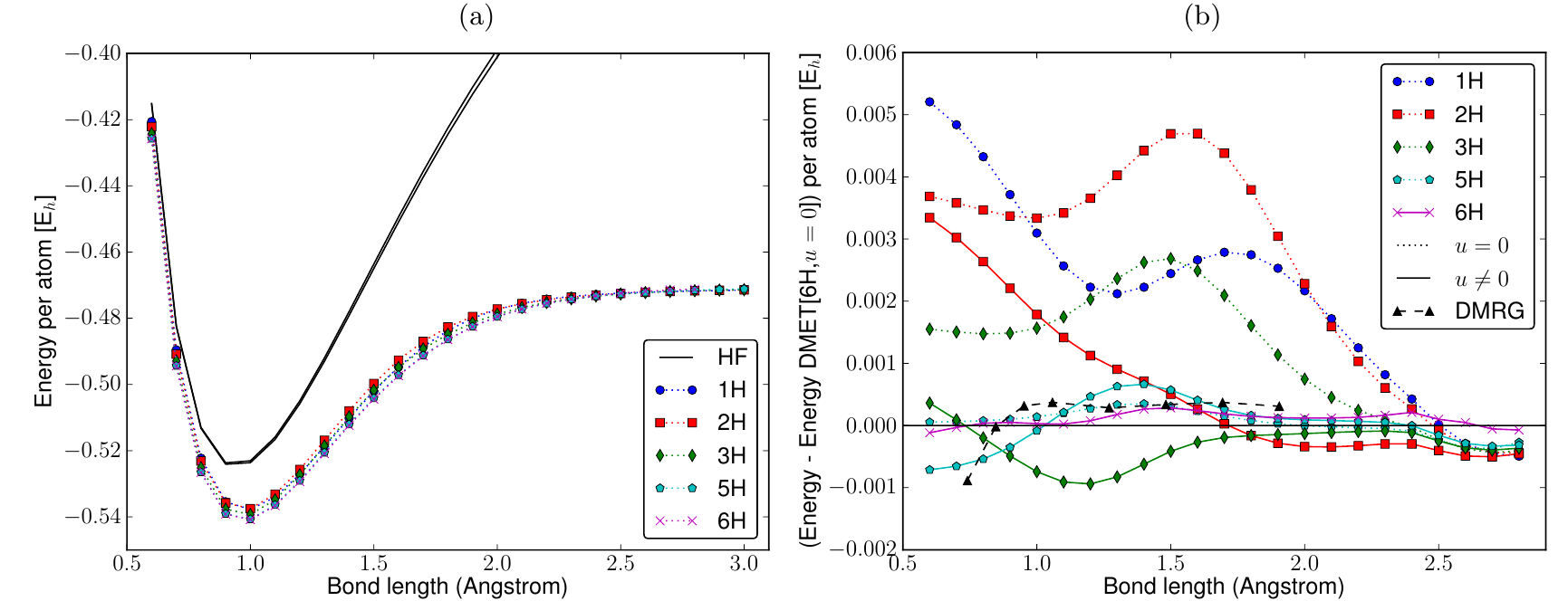}
 \caption{\label{hchain_figure2} Interacting bath DMET results for the
   symmetric stretch of a hydrogen ring with 90 atoms in the STO-6G
   basis (using a L\"owdin symmetric orthogonalization). (a) Bond
   dissociation curve. (b) Energy differences with respect to DMET(6H,
   $u=0$) calculations.}
\end{figure}

We show in Fig. \ref{hchain_figure2} interacting bath DMET results for
the symmetric stretch of a 90-atom hydrogen ring using the STO-6G
basis, a L\"owdin symmetric orthogonalization, and FCI as the
high-level solver. In this case, we show results for fragments with
one, two, three, five, and six hydrogen atoms. The fully symmetric RHF
solution was used to carry out the DMET calculations, even though the
dimerized RHF solution yields lower energies across the entire
dissociation profile. The left plot shows results without a
correlation potential; results appear to converge relatively
monotonically with respect to the size of the fragment around the
equilibrium region. The right plot shows the difference with respect
to the DMET(6H, $u=0$) energies, additionally including results with
$u \neq 0$. We further compare with the DMRG calculations presented in
Ref.~\citenum{Hachmann} for the energy of the 50-atom H chain. (Note
that for short bond lengths larger differences are expected with
respect to DMRG results due to the difference in finite size effects
in a ring vs a chain.) Both $u = 0$ and $u \neq 0$ results using the
larger fragments are only slightly off (a few tenths of a mE$_h$ per
atom) from DMRG for bond lengths greater than 1.0~Angstrom. For the
smaller fragment sizes, the fragment density matrix fitting results
are in better agreement with DMRG than the single-shot embedding
ones. However, the difference between the two types of
self-consistency becomes small as the fragment size increases.
Unfortunately, convergence with respect to the fragment size here is
non-monotonic, which prevents us from attempting accurate
extrapolations.

\begin{figure}
 \centering
 \includegraphics[trim={2cm 12.3cm 2cm 1.8cm},clip,width=1.0\textwidth]{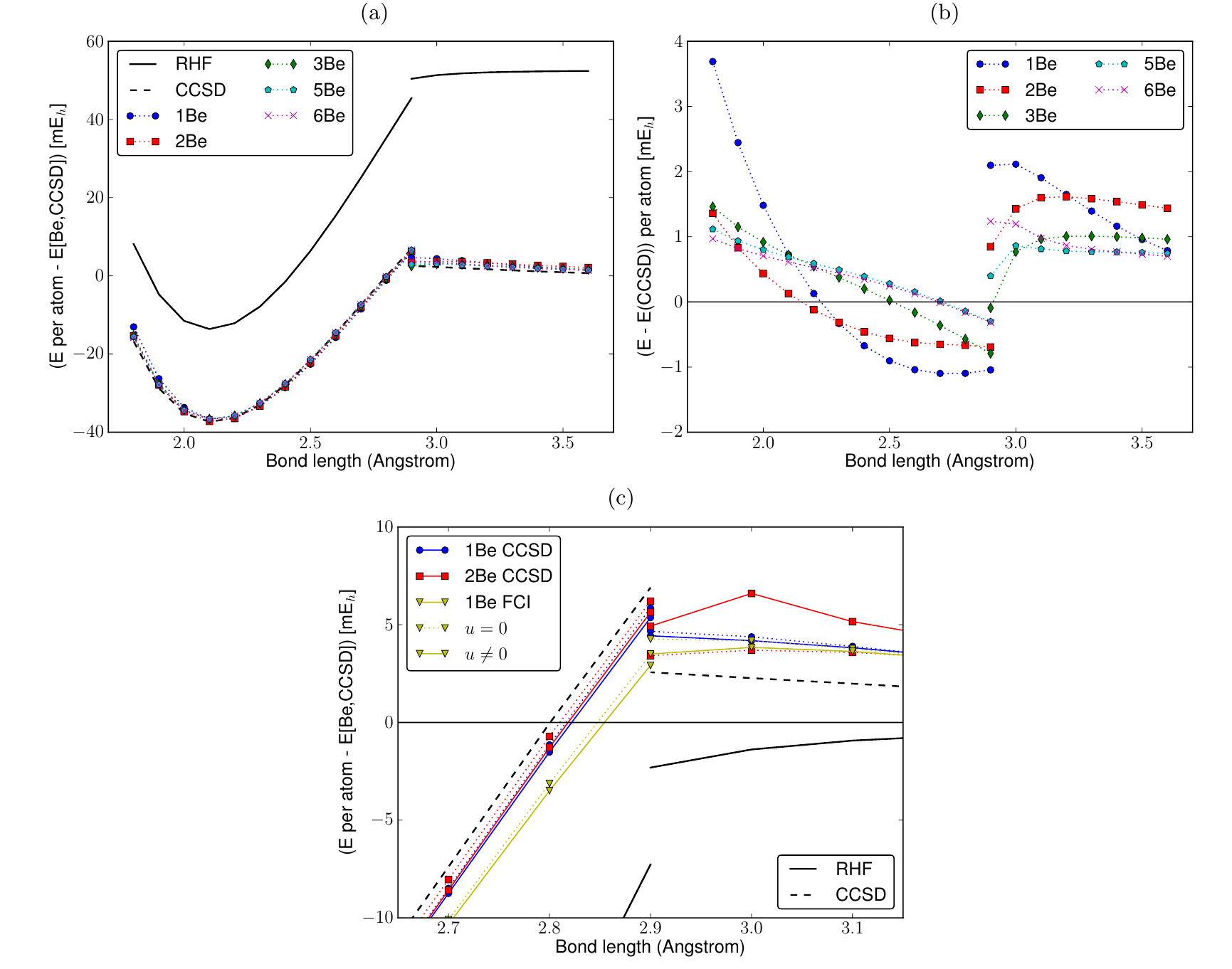}
 \caption{\label{beryllium_figure} Interacting bath DMET results for
   the symmetric stretch of a beryllium ring with 30 atoms using the
   STO-6G basis set. (a) Potential energy curve obtained in DMET $u=0$
   calculations (using a CCSD solver) compared to CCSD and RHF. (b)
   Differences in DMET $u=0$ calculations compared to full-system
   CCSD. (c) Zoom-in of panel (a) near the curve-crossing region,
   additionally comparing with $u \neq 0$ results and with the use of
   a FCI solver. The RHF energies in (c) have been shifted by the
   atomic CCSD correlation energy.}
\end{figure}

\subsection{Beryllium rings}

In this section, we consider results for a ring of 30 beryllium atoms
using the STO-6G basis set. As shown in Fig.~\ref{beryllium_figure},
the RHF solutions near equilibrium and towards dissociation have
different character. In the former case, there is $\sigma$ ($sp$-$sp$)
bonding between the beryllium atoms, while in the latter case the
atomic $2s$ orbitals are occupied.\cite{paulus}

The DMET calculations used restricted HF as the low-level method and
either FCI or CCSD as the high-level method. Our localization
procedure for DMET calculations proceeded as follows. We first defined
the core orbitals by projecting the RHF occupied MOs into atomic-like
$1s$ orbitals. We then obtained IAOs using the atomic-like $2s$ and a
single $2p$ orbital (tangent to the ring). The remaining $2p$ orbitals
were used to define the virtual space. This localization strategy
leads to a $(7,6)$ active space when a single beryllium atom is used
to define the fragment. Self-consistent DMET calculations were
performed in two stages: first, the core and virtual orbitals were
treated as frozen and a correlation potential was optimized in DMET
calculations using the $2s$ and (tangent) $2p$ orbitals with the cost
function Eq.~\eqref{costfunction2_improved}; the correlation potential
thus obtained was included later in a single-shot embedding
calculation using the full set of orbitals.

We have not computed exact benchmark data for this system. Instead, we
have compared our DMET energies to the full system CCSD energies. When
the correlation is not too strong, we expect the full system CCSD to
be an accurate benchmark. However, under more strongly correlated
conditions, for example, as the bonds are stretched, or near an
avoided crossing, we might expect small fragment DMET calculations
with a CCSD solver\footnote{Note that it is often the case that the
  impurity problem is more amenable to a many-body correlation
  treatment than the original problem. In particular, if there is no
  bath truncation and the highest occupied MO (HOMO) or the lowest
  unoccupied MO (LUMO) are delocalized over the fragment and the
  environment, it follows that the single-particle gap is larger on
  the impurity.}  to be more accurate than the full system CCSD
itself, as the latter can break down.

As shown in Fig. \ref{beryllium_figure} (a), the single-shot DMET
energies generally lie close to the full system CCSD results along the
whole dissociation curve. All curves are discontinuous, due to the
crossing of the two RHF solutions with different character.  If we
examine the difference from the full system CCSD in
Fig. \ref{beryllium_figure} (b), the largest difference (with a 1-atom
fragment) is smaller than 4 mE$_h$ per atom, while larger fragments (5
or 6 Be atoms) give differences of only $\approx$ 1 mE$_h$ per atom
along the entire dissociation profile. Close to, and to the right of,
the RHF crossing point we see the largest differences of the DMET
energies from the full system CCSD energy. This deviation does not
significantly decrease with the larger fragment sizes. The explanation
is found in Fig. \ref{beryllium_figure} (c), which provides a close-up
of the energies around the crossing region. We see that the CCSD
energies display an unphysical discontinuous jump comparable to that
of the RHF solution. However, the DMET energies have a much smaller
jump, much closer to the correct physical result.  The size of this
jump is much smaller with self-consistency, indicating that the DMET
self-consistency can remove most of the dependence on the initial RHF
determinant. Indeed, for the points near the crossing we see a smooth
transition in the character of the DMET low-level wavefunction from
doubly occupied $2s$ to $sp$ hybridization. This thus illustrates a
situation where small fragment DMET using an approximate high-level
method yields better behaviour than a calculation on the full system.
Indeed, the difference between using FCI or CCSD as a solver in these
DMET fragment sizes appears quite small, which would not be the case
for the full system.

\subsection{S$_{\text{N}}$2 reaction} \label{sn2_reaction_section}
\begin{figure}
 \centering
 \includegraphics[width=0.5\textwidth]{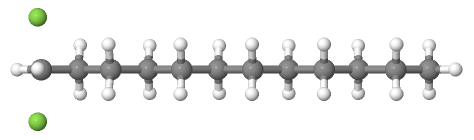}
 \caption{\label{ts_figure} Optimized transition state geometry for the S$_{\text{N}}$2 reaction.}
\end{figure}

In this section, we study single-shot embedding (ie., $u = 0$ but with
a global chemical potential) DMET results with an interacting bath
Hamiltonian for the symmetric S$_{\text{N}}$2 reaction
\begin{equation} 
  \text{C}_{12}\text{H}_{25}\text{F} \cdots \text{F}^{-} \longrightarrow \text{C}_{12}\text{H}_{25}\text{F} \cdots \text{F}^{-}. \label{sn2_reaction_equation}
\end{equation}
The transition state and geometries along the internal reaction
coordinate (IRC) were optimized with \textsc{gaussian09}\citep{g09}
and the B3LYP method, along with the cc-pVTZ basis~\cite{ccbasis} for
C and H and the aug-cc-pVTZ basis~\cite{augccbasis} for F atoms. In
the interacting bath DMET calculations we used the cc-pVDZ
basis~\cite{ccbasis} for all atoms. The transition state is shown in
Fig. \ref{ts_figure}.

In the DMET calculations presented below, we have used the IAO-based
localization procedure described in
Sec.~\ref{practicalbathorbitalconstructionsection}. The system was
partitioned into fragments by cutting across C-C bonds. Different
fragment sizes, labeled by the number of carbon atoms in each fragment
({\#}C), were considered. If 1C fragments are used, then the leftmost
fragment corresponds to a CH$_2$F$_2$ unit, followed by a CH$_2$ unit,
and so on. Restricted HF was used as the low-level method. Three
types of calculations have been performed:
\begin{enumerate}
  \item A standard DMET calculation using CCSD as the high-level
    method for each fragment, denoted as DMET(all).
  \item A DMET calculation where only the leftmost fragment (where the
    substitution takes place) is treated with the high-level method
    (CCSD), while others are treated at the RHF level.\footnote{Here,
      RHF is used as a high-level method to solve each impurity
      Hamiltonian. The resulting 1-RDM may differ slightly from that
      of the original Slater determinant $\ket{\Phi}$ due to
      truncation of the bath orbital space and the presence of
      $\mu_{\text{glob}}$.}  This we label as DMET(1).
  \item Same as above, but with the active space formula for the
    energy (Eq.~\ref{active_space_energy}) and particle number,
    denoted AS.
\end{enumerate}
Note that for $E_{\text{AS}}$ the global particle number is
automatically correct, but the global chemical potential
$\mu_{\text{glob}}$ needs to be optimized for the former two cases.
In selecting the bath orbitals for a given fragment, we have
considered two different schemes: truncating the space using an
eigenvalue cutoff of $10^{-13}$, and keeping a single bath orbital per
chemical bond broken.\cite{qimingJCTC}

\begin{figure}
 \centering
 \includegraphics[height=0.36\textwidth]{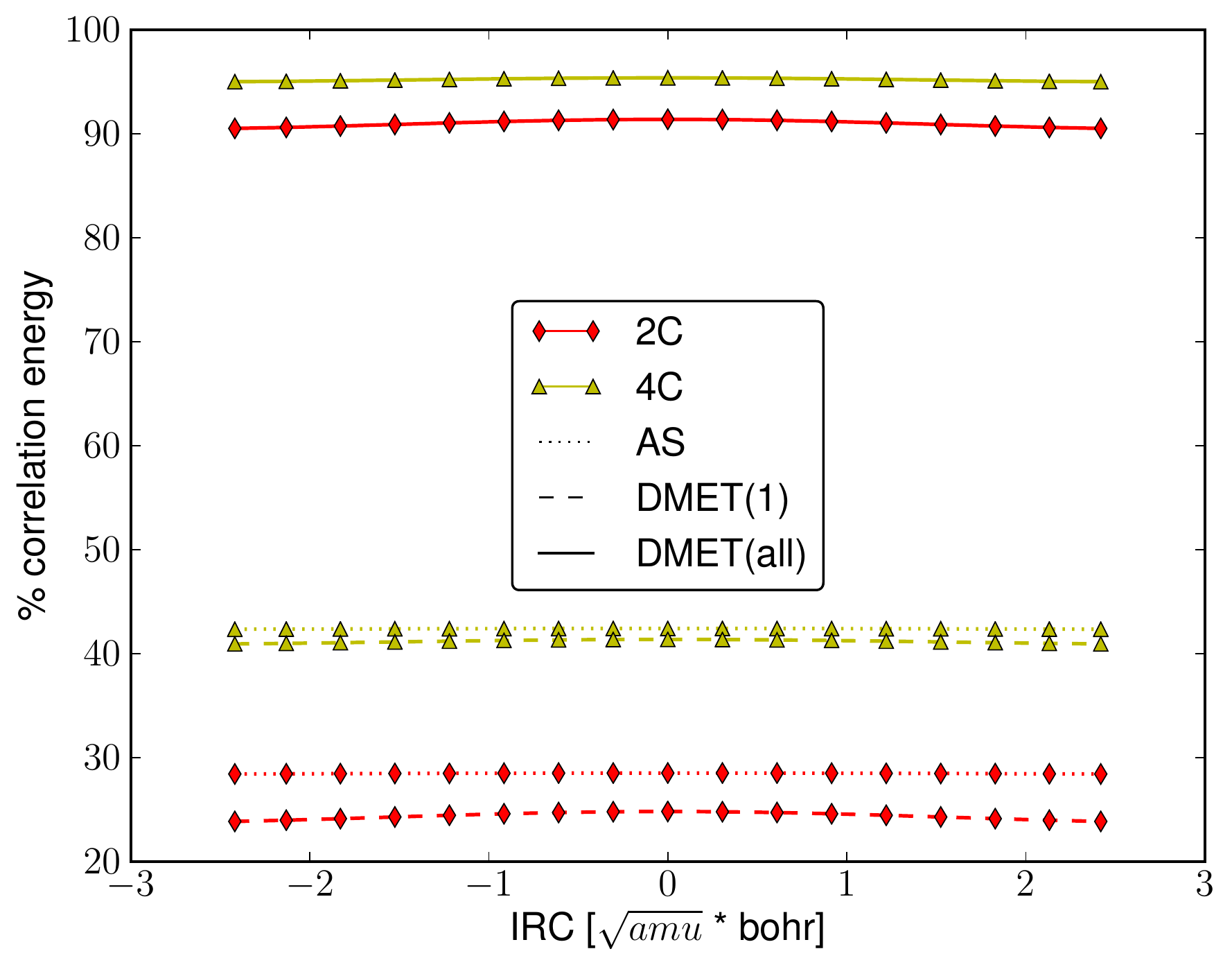}
 \caption{\label{sn2_total_corr} Fraction of correlation energy
   obtained in single-shot DMET(1), DMET(all), and AS calculations
   using 2C and 4C fragments along the IRC of the S$_{\text{N}}$2
   reaction \eqref{sn2_reaction_equation}. Here, the bath orbital
   space includes a single orbital per chemical bond cut.}
\end{figure}

Fig.~\ref{sn2_total_corr} shows the fraction of correlation energy
(with respect to full-system CCSD) obtained with the different
calculation schemes. It is clear that the total energies from
DMET(all) calculations are more accurate than with other schemes, as
correlation from all electrons is accounted for. Nevertheless, the
same is not true for the relative energy profiles, as discussed below.

\begin{figure}
 \centering
 \includegraphics[trim={2cm 5.4cm 2cm 1.8cm},clip,width=1.0\textwidth]{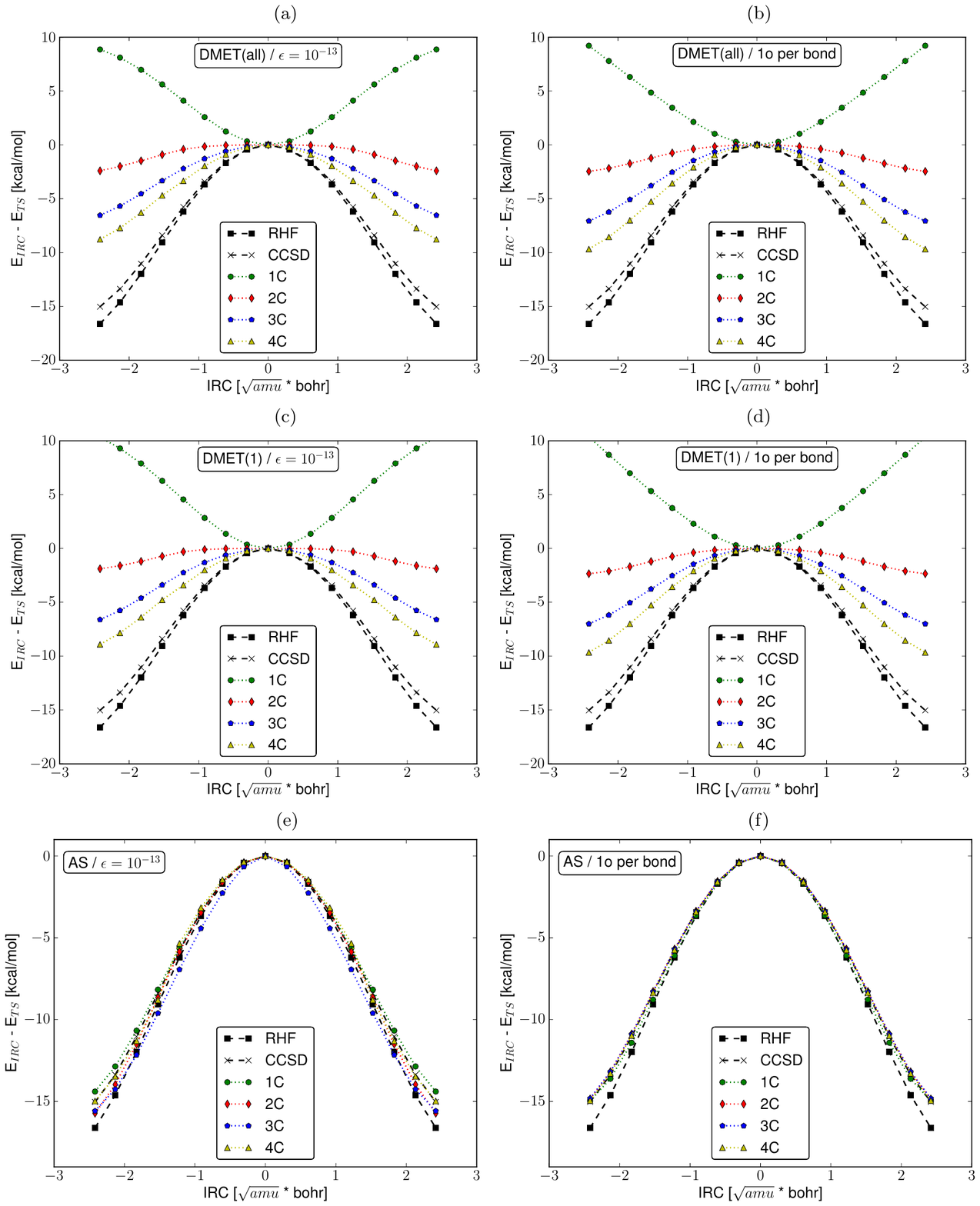}
 \caption{\label{sn2_iao_all_figure} Single-shot interacting bath
   DMET(1), DMET(all), and AS relative energies in the IAO-localized
   cc-pVDZ basis set for the S$_{\text{N}}$2 reaction
   \eqref{sn2_reaction_equation}, using either the occupation number
   cutoff $\epsilon=10^{-13}$ to select bath orbitals or selecting one
   bath orbital per chemical bond cut. The text box in the top of
   figures (a)-(f) indicates the combination of energy formula and
   bath selection.}
\end{figure}

Fig.~\ref{sn2_iao_all_figure} displays the relative energy profiles:
results using an eigenvalue threshold of $\epsilon=10^{-13}$ to select
bath orbitals are shown in panels (a),(c),(e), while those with a
single bath orbital per chemical bond cut are shown in panels
(b),(d),(f). Although the behavior shown in (a) and (c) significantly
improves with increasing fragment size, the qualitative behavior is
still far from the CCSD or RHF one. On the other hand, the active
space energy (e) displays a qualitatively correct behavior even with
1C fragments, while approaching quantitative agreement with CCSD as
the fragment size is increased. The very permissive threshold of
$\epsilon=10^{-13}$ captures a large amount of bath orbitals per
fragment, and this is partly responsible for the slow convergence of
the AS energy with respect to CCSD. It is even responsible for a small
jump in the 4C energy profile at $|IRC| \approx 1.5 \,
\sqrt{amu}$~bohr (hard to see), as an additional bath orbital is
included for larger values of IRC. If the bath selection is restricted
to 1 orbital per chemical bond cut (see panels (b),(d),(f)), the
agreement with CCSD observed in AS energy calculations is much
better. The DMET(all) and DMET(1) profiles are not significantly
changed.

\begin{figure}
 \centering
 \includegraphics[trim={2cm 19.4cm 2cm 1.8cm},clip,width=1.0\textwidth]{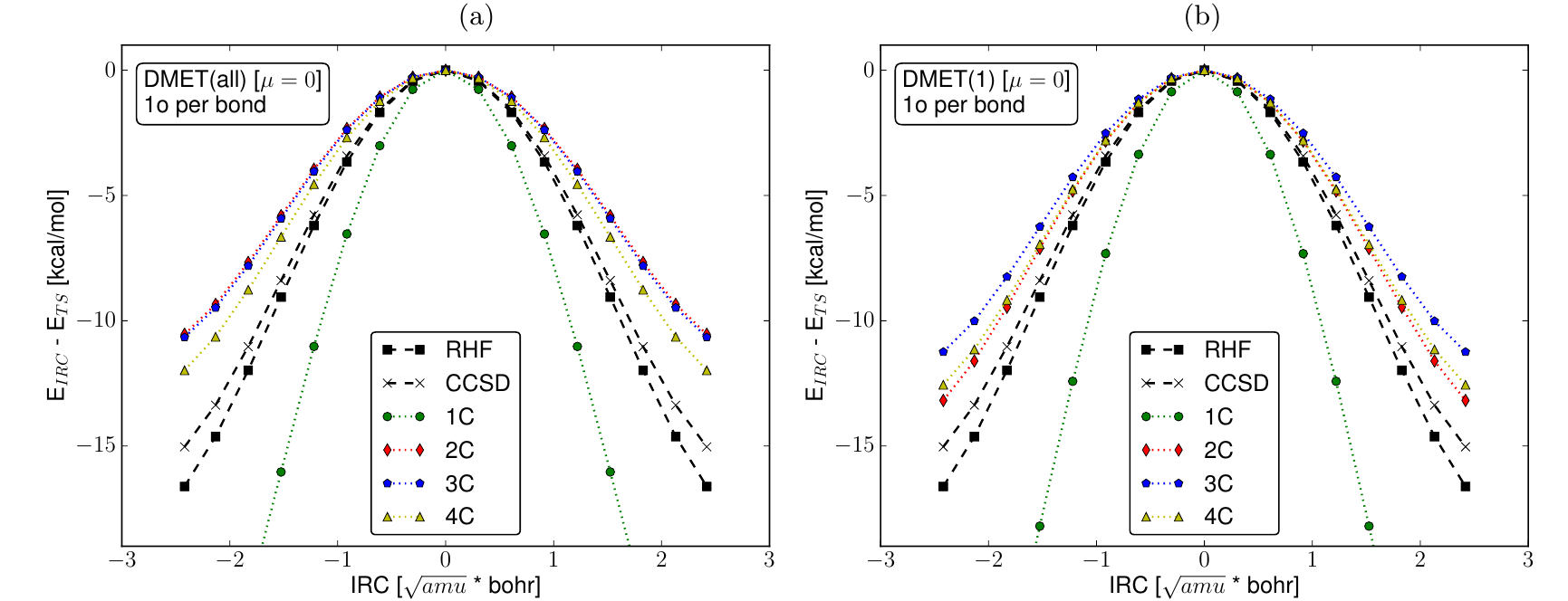}
 \caption{\label{fig:sn2_mu0} Same as
   Fig.~\ref{sn2_iao_all_figure}. $\mu_{\text{glob}} = 0$ DMET(1) and
   DMET(all) relative energies using one bath orbital per chemical
   bond cut.}
\end{figure}

We can analyze the origin of the poor behavior of the DMET(all) and
DMET(1) schemes. In particular, the only difference between (c) and
(e) (or, equivalently, (d) and (f)) is that (c) uses the standard DMET
democratic partitioning of the expectation values across fragments for
both the particle number and energy.  Although this democratic
partitioning allows information from different fragments to contribute
equally to the whole calculation (important, for example, in a
translationally invariant system) this is not advantageous in the
current example as the chemical change is occurring purely locally. In
Fig.~\ref{fig:sn2_mu0} we further show the energy profile
corresponding to (b), (d) using the standard DMET formula for the
energy, but without adjusting the global chemical potential. In this
case, the energy profile appears improved and qualitatively correct,
although the convergence with respect to fragment size is still
slow. This indicates that it is the democratic evaluation of the
particle number which yields the largest source of error in the DMET
calculations.

\section{Summary} \label{section_summary}

In this work we have reviewed several aspects of density matrix
embedding theory (DMET) in detail. In
Sec. \ref{section_bath_orbital_construction}, we discuss how a bath
space can be constructed for a local fragment. While correlated
low-level methods provide accurate many-body bath states, most quantum
chemistry solvers are formulated in terms of orbitals. The Schmidt
decomposition of a mean-field wavefunction naturally gives rise to
bath orbitals. We have reviewed the DMET bath orbital construction,
and provided a practical way to obtain the bath orbitals from the
mean-field 1-RDM of the total system. In the future, it will be
interesting to construct a bath \textit{orbital} space from a
correlated low-level wavefunction.

In Sec. \ref{active_space_ahm_major_Section}, we discuss the
construction of both the non-interacting bath and interacting bath
low-level and high-level Hamiltonians. Once the high-level Hamiltonian
problem is solved, and the corresponding 1- and 2-RDMs are obtained in
the local fragment and bath orbital spaces, DMET energies can be
calculated based on the formulae in
Sec. \ref{DMETenergy_major_seciton}. In order to fine-tune the
low-level wavefunction to construct better bath spaces, a DMET
correlation potential is introduced. Its self-consistent optimization
is discussed in
Sec. \ref{correlationpotentialoptimization_subsection}, as well as the
optimization of a global chemical potential shift to ensure that the
fragments contain the correct total number of electrons. This section
also provides an overview of how the different parts of DMET fit into
the full DMET algorithm.

In Sec. \ref{section_applications} several applications are
studied. In hydrogen and beryllium rings we consider calculations with
a single-shot embedding scheme and in a full self-consistent
correlation potential DMET treatment. The effect of self-consistency
is generally minor, but becomes pronounced near drastic changes in the
character of the HF solution, where the optimal DMET Slater
determinant may differ considerably from the HF one. In hydrogen and
beryllium rings our DMET calculations have nearly quantitative
agreement with accurate dissociation profiles even when small impurity
sizes are used. The agreement improves significantly as the size of
the impurity is increased. In the beryllium rings, self-consistency is
important for describing the avoided crossing region, and the DMET
calculations with small fragments, using an approximate coupled
cluster solver, appear more accurate than the full system coupled
cluster results themselves.

For the reaction barrier of an S$_{\text{N}}$2 reaction, we have
tested single-shot active space energies with CCSD as an active space
solver, the DMET energy formula where only one impurity is treated
with CCSD as the high-level method, and the DMET energy formula where
all impurities are treated with CCSD as the high-level method. In
addition, we compare the accuracy of a large cutoff-based bath orbital
space with the selection of one bath orbital per chemical bond cut. We
have found that the active space {\em relative} energies converge the
fastest to the CCSD calculations for the full system. This is because
the standard DMET democratic evaluation of expectation values across
fragments does not provide optimal error cancellation when only local
changes in a single fragment take place.  Thus, for molecular
applications, when reactions occurs locally, the single-shot DMET
active space energies with one bath orbital per chemical bond cut
provide the most reliable description. Note that if FCI is used as an
active space solver, this exactly corresponds to a CAS-CI calculation,
with DMET providing a natural way to define the {\em relevant} active
space.

\begin{acknowledgement}
  S.W. gratefully acknowledges a Gustave Bo\"el - Sofina -
  B.A.E.F. postdoctoral fellowship from the King Baudouin Foundation
  and the Belgian-American Educational Foundation for the academic
  year 2014-2015. G. K.-L. C. acknowledges support from the US
  Department of Energy through DE-SC0010530. Additional support was
  provided from the Simons Foundation through the Simons Collaboration
  on the Many-Electron Problem.
\end{acknowledgement}

\appendix
\section{Analytic gradients of the mean-field density matrix with respect to the correlation potential\label{appendixGRAD}}
Consider a change $\delta$ in one particular value of the correlation
potential. The change in the mean-field operator can be written as:
\begin{equation}
\hat{H} = \hat{H}^0 + \delta \hat{H}^1,
\end{equation}
where both $\hat{H}^0$ and $\hat{H}^1$ are Hermitian $L \times L$
matrices. With the mean-field solution:
\begin{equation}
\hat{H}^0 = \left[ \begin{array}{cc} C_{\text{occ}}^0 & C_{\text{vir}}^0 \end{array} \right] \left[ \begin{array}{cc} E^0_{\text{occ}} & \\ & E^0_{\text{vir}} \end{array} \right] \left[ \begin{array}{c} C_{\text{occ}}^{0,\dagger} \\ C_{\text{vir}}^{0,\dagger} \end{array} \right],
\end{equation}
one can solve for the first order (Rayleigh-Schr\"odinger) response
equation:
\begin{equation}
\hat{H}^0 C^1_{\text{occ}} + \hat{H}^1 C^0_{\text{occ}} = C^1_{\text{occ}} E^0_{\text{occ}} + C^0_{\text{occ}} E^1_{\text{occ}}. \label{MFresponseEq1}
\end{equation}
The matrices $C_{\text{occ}}^i$ have the shape $L \times
N_{\text{occ}}$ and represent the order $i$ occupied orbitals. The
matrices $C_{\text{vir}}^i$ have the shape $L \times (L -
N_{\text{occ}})$ and represent the order $i$ virtual orbitals. The
diagonal matrix $E^0_{\text{occ}}$ has the shape $N_{\text{occ}}
\times N_{\text{occ}}$ and represents the occupied orbital energies,
and likewise for $E^0_{\text{vir}}$. The occupied first order response
orbitals are orthogonal to the ground-state orbitals:
\begin{equation}
C^{0,\dagger}_{\text{occ}} C^{1}_{\text{occ}} = 0 \quad \Rightarrow \quad E^1_{\text{occ}} = C^{0,\dagger}_{\text{occ}} \hat{H}^1 C^{0}_{\text{occ}}. \label{orthoresponseEQ}
\end{equation}
This allows to rewrite Eq. \eqref{MFresponseEq1} as
\begin{equation}
\hat{H}^0 C^1_{\text{occ}} - C^1_{\text{occ}} E^0_{\text{occ}} = -(1 - C^{0}_{\text{occ}} C^{0,\dagger}_{\text{occ}} ) \hat{H}^1 C^0_{\text{occ}} = - C^{0}_{\text{vir}} C^{0,\dagger}_{\text{vir}} \hat{H}^1 C^0_{\text{occ}}. \label{MFresponseEq2}
\end{equation}
By virtue of Eq. \eqref{orthoresponseEQ}, the response orbitals can be written as
\begin{equation}
C^1_{\text{occ}} = C^0_{\text{vir}} Z^1.
\end{equation}
The entries of the $(L - N_{\text{occ}}) \times N_{\text{occ}}$ matrix
$Z^1$ can be found with Eq. \eqref{MFresponseEq2}:
\begin{equation}
Z^1_{\mu \nu} = - \frac{ \left( C^{0,\dagger}_{\text{vir}} \hat{H}^1 C^0_{\text{occ}} \right)_{\mu \nu}}{E^0_{\text{vir}, \mu} - E^0_{\text{occ}, \nu}}.
\end{equation}
Finally, the first order response of the density matrix can be
obtained as:
\begin{eqnarray}
\left. \frac{\partial D}{\partial \delta} \right|_{\delta=0} & = & \left. \frac{\partial}{\partial \delta} \left[ (C^0_{\text{occ}} + \delta C^1_{\text{occ}}) (C^0_{\text{occ}} + \delta C^1_{\text{occ}})^{\dagger} \right] \right|_{\delta=0} \nonumber \\
& = & C^0_{\text{occ}} C^{1,\dagger}_{\text{occ}} + C^1_{\text{occ}} C^{0,\dagger}_{\text{occ}} = C^0_{\text{occ}} Z^{1,\dagger} C^{0,\dagger}_{\text{vir}} + C^0_{\text{vir}} Z^1 C^{0,\dagger}_{\text{occ}}.
\end{eqnarray}


\begin{mcitethebibliography}{42}
\providecommand*\natexlab[1]{#1}
\providecommand*\mciteSetBstSublistMode[1]{}
\providecommand*\mciteSetBstMaxWidthForm[2]{}
\providecommand*\mciteBstWouldAddEndPuncttrue
  {\def\EndOfBibitem{\unskip.}}
\providecommand*\mciteBstWouldAddEndPunctfalse
  {\let\EndOfBibitem\relax}
\providecommand*\mciteSetBstMidEndSepPunct[3]{}
\providecommand*\mciteSetBstSublistLabelBeginEnd[3]{}
\providecommand*\EndOfBibitem{}
\mciteSetBstSublistMode{f}
\mciteSetBstMaxWidthForm{subitem}{(\alph{mcitesubitemcount})}
\mciteSetBstSublistLabelBeginEnd
  {\mcitemaxwidthsubitemform\space}
  {\relax}
  {\relax}

\bibitem[Coester and K\"ummel(1960)Coester, and K\"ummel]{Coester1960477}
Coester,~F.; K\"ummel,~H. \emph{Nucl. Phys.} \textbf{1960}, \emph{17},
  477--485\relax
\mciteBstWouldAddEndPuncttrue
\mciteSetBstMidEndSepPunct{\mcitedefaultmidpunct}
{\mcitedefaultendpunct}{\mcitedefaultseppunct}\relax
\EndOfBibitem
\bibitem[Cizek(1966)]{CCcizek}
Cizek,~J. \emph{J. Chem. Phys.} \textbf{1966}, \emph{45}, 4256--4266\relax
\mciteBstWouldAddEndPuncttrue
\mciteSetBstMidEndSepPunct{\mcitedefaultmidpunct}
{\mcitedefaultendpunct}{\mcitedefaultseppunct}\relax
\EndOfBibitem
\bibitem[Roger and Hetherington(1990)Roger, and Hetherington]{CClattice}
Roger,~M.; Hetherington,~J.~H. \emph{Europhys. Lett.} \textbf{1990}, \emph{11},
  255\relax
\mciteBstWouldAddEndPuncttrue
\mciteSetBstMidEndSepPunct{\mcitedefaultmidpunct}
{\mcitedefaultendpunct}{\mcitedefaultseppunct}\relax
\EndOfBibitem
\bibitem[White(1992)]{PhysRevLett.69.2863}
White,~S.~R. \emph{Phys. Rev. Lett.} \textbf{1992}, \emph{69}, 2863--2866\relax
\mciteBstWouldAddEndPuncttrue
\mciteSetBstMidEndSepPunct{\mcitedefaultmidpunct}
{\mcitedefaultendpunct}{\mcitedefaultseppunct}\relax
\EndOfBibitem
\bibitem[White and Martin(1999)White, and Martin]{whiteQC}
White,~S.~R.; Martin,~R.~L. \emph{J. Chem. Phys.} \textbf{1999}, \emph{110},
  4127--4130\relax
\mciteBstWouldAddEndPuncttrue
\mciteSetBstMidEndSepPunct{\mcitedefaultmidpunct}
{\mcitedefaultendpunct}{\mcitedefaultseppunct}\relax
\EndOfBibitem
\bibitem[Dukelsky and Pittel(2001)Dukelsky, and Pittel]{PhysRevC.63.061303}
Dukelsky,~J.; Pittel,~S. \emph{Phys. Rev. C} \textbf{2001}, \emph{63},
  061303\relax
\mciteBstWouldAddEndPuncttrue
\mciteSetBstMidEndSepPunct{\mcitedefaultmidpunct}
{\mcitedefaultendpunct}{\mcitedefaultseppunct}\relax
\EndOfBibitem
\bibitem[Metzner and Vollhardt(1989)Metzner, and Vollhardt]{PhysRevLett.62.324}
Metzner,~W.; Vollhardt,~D. \emph{Phys. Rev. Lett.} \textbf{1989}, \emph{62},
  324--327\relax
\mciteBstWouldAddEndPuncttrue
\mciteSetBstMidEndSepPunct{\mcitedefaultmidpunct}
{\mcitedefaultendpunct}{\mcitedefaultseppunct}\relax
\EndOfBibitem
\bibitem[Georges and Krauth(1992)Georges, and Krauth]{PhysRevLett.69.1240}
Georges,~A.; Krauth,~W. \emph{Phys. Rev. Lett.} \textbf{1992}, \emph{69},
  1240--1243\relax
\mciteBstWouldAddEndPuncttrue
\mciteSetBstMidEndSepPunct{\mcitedefaultmidpunct}
{\mcitedefaultendpunct}{\mcitedefaultseppunct}\relax
\EndOfBibitem
\bibitem[Zgid and Chan(2011)Zgid, and Chan]{zgidDMFT}
Zgid,~D.; Chan,~G. K.-L. \emph{J. Chem. Phys.} \textbf{2011}, \emph{134},
  094115\relax
\mciteBstWouldAddEndPuncttrue
\mciteSetBstMidEndSepPunct{\mcitedefaultmidpunct}
{\mcitedefaultendpunct}{\mcitedefaultseppunct}\relax
\EndOfBibitem
\bibitem[Lin \latin{et~al.}(2011)Lin, Marianetti, Millis, and
  Reichman]{PhysRevLett.106.096402}
Lin,~N.; Marianetti,~C.~A.; Millis,~A.~J.; Reichman,~D.~R. \emph{Phys. Rev.
  Lett.} \textbf{2011}, \emph{106}, 096402\relax
\mciteBstWouldAddEndPuncttrue
\mciteSetBstMidEndSepPunct{\mcitedefaultmidpunct}
{\mcitedefaultendpunct}{\mcitedefaultseppunct}\relax
\EndOfBibitem
\bibitem[Knizia and Chan(2012)Knizia, and Chan]{geraldPRL}
Knizia,~G.; Chan,~G. K.-L. \emph{Phys. Rev. Lett.} \textbf{2012}, \emph{109},
  186404\relax
\mciteBstWouldAddEndPuncttrue
\mciteSetBstMidEndSepPunct{\mcitedefaultmidpunct}
{\mcitedefaultendpunct}{\mcitedefaultseppunct}\relax
\EndOfBibitem
\bibitem[Knizia and Chan(2013)Knizia, and Chan]{geraldJCTC}
Knizia,~G.; Chan,~G. K.-L. \emph{J. Chem. Theory Comput.} \textbf{2013},
  \emph{9}, 1428--1432\relax
\mciteBstWouldAddEndPuncttrue
\mciteSetBstMidEndSepPunct{\mcitedefaultmidpunct}
{\mcitedefaultendpunct}{\mcitedefaultseppunct}\relax
\EndOfBibitem
\bibitem[Wouters and Van~Neck(2014)Wouters, and Van~Neck]{EPJDreview}
Wouters,~S.; Van~Neck,~D. \emph{Eur. Phys. J. D} \textbf{2014}, \emph{68},
  272\relax
\mciteBstWouldAddEndPuncttrue
\mciteSetBstMidEndSepPunct{\mcitedefaultmidpunct}
{\mcitedefaultendpunct}{\mcitedefaultseppunct}\relax
\EndOfBibitem
\bibitem[Georges \latin{et~al.}(1996)Georges, Kotliar, Krauth, and
  Rozenberg]{RevModPhys.68.13}
Georges,~A.; Kotliar,~G.; Krauth,~W.; Rozenberg,~M.~J. \emph{Rev. Mod. Phys.}
  \textbf{1996}, \emph{68}, 13--125\relax
\mciteBstWouldAddEndPuncttrue
\mciteSetBstMidEndSepPunct{\mcitedefaultmidpunct}
{\mcitedefaultendpunct}{\mcitedefaultseppunct}\relax
\EndOfBibitem
\bibitem[Zheng and Chan(2016)Zheng, and Chan]{boxiaoHubbard}
Zheng,~B.-X.; Chan,~G. K.-L. \emph{Phys. Rev. B} \textbf{2016}, \emph{93},
  035126\relax
\mciteBstWouldAddEndPuncttrue
\mciteSetBstMidEndSepPunct{\mcitedefaultmidpunct}
{\mcitedefaultendpunct}{\mcitedefaultseppunct}\relax
\EndOfBibitem
\bibitem[Tsuchimochi \latin{et~al.}(2015)Tsuchimochi, Welborn, and
  Van~Voorhis]{troyGeminals}
Tsuchimochi,~T.; Welborn,~M.; Van~Voorhis,~T. \emph{J. Chem. Phys.}
  \textbf{2015}, \emph{143}, 024107\relax
\mciteBstWouldAddEndPuncttrue
\mciteSetBstMidEndSepPunct{\mcitedefaultmidpunct}
{\mcitedefaultendpunct}{\mcitedefaultseppunct}\relax
\EndOfBibitem
\bibitem[Sandhoefer and Chan(2016)Sandhoefer, and Chan]{barbara}
Sandhoefer,~B.; Chan,~G. K.-L. \emph{arXiv 1602.04195} \textbf{2016}, \relax
\mciteBstWouldAddEndPunctfalse
\mciteSetBstMidEndSepPunct{\mcitedefaultmidpunct}
{}{\mcitedefaultseppunct}\relax
\EndOfBibitem
\bibitem[Fan and Jie(2015)Fan, and Jie]{spinsystemPRB}
Fan,~Z.; Jie,~Q.-L. \emph{Phys. Rev. B} \textbf{2015}, \emph{91}, 195118\relax
\mciteBstWouldAddEndPuncttrue
\mciteSetBstMidEndSepPunct{\mcitedefaultmidpunct}
{\mcitedefaultendpunct}{\mcitedefaultseppunct}\relax
\EndOfBibitem
\bibitem[Chen \latin{et~al.}(2014)Chen, Booth, Sharma, Knizia, and
  Chan]{qiaoniPRB}
Chen,~Q.; Booth,~G.~H.; Sharma,~S.; Knizia,~G.; Chan,~G. K.-L. \emph{Phys. Rev.
  B} \textbf{2014}, \emph{89}, 165134\relax
\mciteBstWouldAddEndPuncttrue
\mciteSetBstMidEndSepPunct{\mcitedefaultmidpunct}
{\mcitedefaultendpunct}{\mcitedefaultseppunct}\relax
\EndOfBibitem
\bibitem[Bulik \latin{et~al.}(2014)Bulik, Chen, and Scuseria]{bulikJCP}
Bulik,~I.~W.; Chen,~W.; Scuseria,~G.~E. \emph{J. Chem. Phys.} \textbf{2014},
  \emph{141}, 054113\relax
\mciteBstWouldAddEndPuncttrue
\mciteSetBstMidEndSepPunct{\mcitedefaultmidpunct}
{\mcitedefaultendpunct}{\mcitedefaultseppunct}\relax
\EndOfBibitem
\bibitem[Bulik \latin{et~al.}(2014)Bulik, Scuseria, and Dukelsky]{bulikPRB}
Bulik,~I.~W.; Scuseria,~G.~E.; Dukelsky,~J. \emph{Phys. Rev. B} \textbf{2014},
  \emph{89}, 035140\relax
\mciteBstWouldAddEndPuncttrue
\mciteSetBstMidEndSepPunct{\mcitedefaultmidpunct}
{\mcitedefaultendpunct}{\mcitedefaultseppunct}\relax
\EndOfBibitem
\bibitem[LeBlanc \latin{et~al.}(2015)LeBlanc, Antipov, Becca, Bulik, Chan,
  Chung, Deng, Ferrero, Henderson, Jim\'enez-Hoyos, Kozik, Liu, Millis,
  Prokof'ev, Qin, Scuseria, Shi, Svistunov, Tocchio, Tupitsyn, White, Zhang,
  Zheng, Zhu, and Gull]{PhysRevX.5.041041}
LeBlanc,~J. P.~F.; Antipov,~A.~E.; Becca,~F.; Bulik,~I.~W.; Chan,~G. K.-L.;
  Chung,~C.-M.; Deng,~Y.; Ferrero,~M.; Henderson,~T.~M.;
  Jim\'enez-Hoyos,~C.~A.; Kozik,~E.; Liu,~X.-W.; Millis,~A.~J.;
  Prokof'ev,~N.~V.; Qin,~M.; Scuseria,~G.~E.; Shi,~H.; Svistunov,~B.~V.;
  Tocchio,~L.~F.; Tupitsyn,~I.~S.; White,~S.~R.; Zhang,~S.; Zheng,~B.-X.;
  Zhu,~Z.; Gull,~E. \emph{Phys. Rev. X} \textbf{2015}, \emph{5}, 041041\relax
\mciteBstWouldAddEndPuncttrue
\mciteSetBstMidEndSepPunct{\mcitedefaultmidpunct}
{\mcitedefaultendpunct}{\mcitedefaultseppunct}\relax
\EndOfBibitem
\bibitem[Sun and Chan(2014)Sun, and Chan]{qimingJCTC}
Sun,~Q.; Chan,~G. K.-L. \emph{J. Chem. Theory Comput.} \textbf{2014},
  \emph{10}, 3784--3790\relax
\mciteBstWouldAddEndPuncttrue
\mciteSetBstMidEndSepPunct{\mcitedefaultmidpunct}
{\mcitedefaultendpunct}{\mcitedefaultseppunct}\relax
\EndOfBibitem
\bibitem[Sorella \latin{et~al.}(2015)Sorella, Devaux, Dagrada, Mazzola, and
  Casula]{geminal_ao_construction}
Sorella,~S.; Devaux,~N.; Dagrada,~M.; Mazzola,~G.; Casula,~M. \emph{J. Chem.
  Phys.} \textbf{2015}, \emph{143}, 244112\relax
\mciteBstWouldAddEndPuncttrue
\mciteSetBstMidEndSepPunct{\mcitedefaultmidpunct}
{\mcitedefaultendpunct}{\mcitedefaultseppunct}\relax
\EndOfBibitem
\bibitem[Booth and Chan(2015)Booth, and Chan]{georgePRB}
Booth,~G.~H.; Chan,~G. K.-L. \emph{Phys. Rev. B} \textbf{2015}, \emph{91},
  155107\relax
\mciteBstWouldAddEndPuncttrue
\mciteSetBstMidEndSepPunct{\mcitedefaultmidpunct}
{\mcitedefaultendpunct}{\mcitedefaultseppunct}\relax
\EndOfBibitem
\bibitem[Wouters(2015)]{github_qcdmet}
Wouters,~S. \textsc{qc-dmet}: a python implementation of density matrix
  embedding theory for ab initio quantum chemistry,
  \url{https://github.com/sebwouters/qc-dmet}. 2015\relax
\mciteBstWouldAddEndPuncttrue
\mciteSetBstMidEndSepPunct{\mcitedefaultmidpunct}
{\mcitedefaultendpunct}{\mcitedefaultseppunct}\relax
\EndOfBibitem
\bibitem[Helgaker \latin{et~al.}(2000)Helgaker, Jorgensen, and Olsen]{helgaker}
Helgaker,~T.; Jorgensen,~P.; Olsen,~J. \emph{Molecular Electronic-Structure
  Theory}, 1st ed.; Wiley, New York, 2000\relax
\mciteBstWouldAddEndPuncttrue
\mciteSetBstMidEndSepPunct{\mcitedefaultmidpunct}
{\mcitedefaultendpunct}{\mcitedefaultseppunct}\relax
\EndOfBibitem
\bibitem[Wouters \latin{et~al.}(2013)Wouters, Nakatani, Van~Neck, and
  Chan]{PhysRevB.88.075122}
Wouters,~S.; Nakatani,~N.; Van~Neck,~D.; Chan,~G. K.-L. \emph{Phys. Rev. B}
  \textbf{2013}, \emph{88}, 075122\relax
\mciteBstWouldAddEndPuncttrue
\mciteSetBstMidEndSepPunct{\mcitedefaultmidpunct}
{\mcitedefaultendpunct}{\mcitedefaultseppunct}\relax
\EndOfBibitem
\bibitem[MacDonald(1933)]{PhysRev.43.830}
MacDonald,~J. K.~L. \emph{Phys. Rev.} \textbf{1933}, \emph{43}, 830--833\relax
\mciteBstWouldAddEndPuncttrue
\mciteSetBstMidEndSepPunct{\mcitedefaultmidpunct}
{\mcitedefaultendpunct}{\mcitedefaultseppunct}\relax
\EndOfBibitem
\bibitem[Knizia(2013)]{iao_gerald}
Knizia,~G. \emph{J. Chem. Theory Comput.} \textbf{2013}, \emph{9},
  4834--4843\relax
\mciteBstWouldAddEndPuncttrue
\mciteSetBstMidEndSepPunct{\mcitedefaultmidpunct}
{\mcitedefaultendpunct}{\mcitedefaultseppunct}\relax
\EndOfBibitem
\bibitem[Wu and Yang(2003)Wu, and Yang]{WuYang}
Wu,~Q.; Yang,~W. \emph{J. Chem. Phys.} \textbf{2003}, \emph{118},
  2498--2509\relax
\mciteBstWouldAddEndPuncttrue
\mciteSetBstMidEndSepPunct{\mcitedefaultmidpunct}
{\mcitedefaultendpunct}{\mcitedefaultseppunct}\relax
\EndOfBibitem
\bibitem[Lieb(1983)]{Lieb}
Lieb,~E.~H. \emph{Int. J. Quantum Chem.} \textbf{1983}, \emph{24},
  243--277\relax
\mciteBstWouldAddEndPuncttrue
\mciteSetBstMidEndSepPunct{\mcitedefaultmidpunct}
{\mcitedefaultendpunct}{\mcitedefaultseppunct}\relax
\EndOfBibitem
\bibitem[Sun(2015)]{github_pyscf}
Sun,~Q. \textsc{pyscf}: python module for quantum chemistry,
  \url{https://github.com/sunqm/pyscf}. 2015\relax
\mciteBstWouldAddEndPuncttrue
\mciteSetBstMidEndSepPunct{\mcitedefaultmidpunct}
{\mcitedefaultendpunct}{\mcitedefaultseppunct}\relax
\EndOfBibitem
\bibitem[Wouters \latin{et~al.}(2014)Wouters, Poelmans, Ayers, and
  Neck]{chemps2_cpc}
Wouters,~S.; Poelmans,~W.; Ayers,~P.~W.; Neck,~D.~V. \emph{Comput. Phys.
  Commun.} \textbf{2014}, \emph{185}, 1501--1514\relax
\mciteBstWouldAddEndPuncttrue
\mciteSetBstMidEndSepPunct{\mcitedefaultmidpunct}
{\mcitedefaultendpunct}{\mcitedefaultseppunct}\relax
\EndOfBibitem
\bibitem[Shavitt and Bartlett(2009)Shavitt, and Bartlett]{bartlett}
Shavitt,~I.; Bartlett,~R.~J. \emph{Many-Body Methods in Chemistry and Physics.
  {MBPT} and Coupled-Cluster Theory}, 1st ed.; Cambridge Molecular Science;
  Cambridge University Press, New York, 2009\relax
\mciteBstWouldAddEndPuncttrue
\mciteSetBstMidEndSepPunct{\mcitedefaultmidpunct}
{\mcitedefaultendpunct}{\mcitedefaultseppunct}\relax
\EndOfBibitem
\bibitem[Gauss and Stanton(1995)Gauss, and Stanton]{ccsdrdm2}
Gauss,~J.; Stanton,~J.~F. \emph{J. Chem. Phys.} \textbf{1995}, \emph{103},
  3561--3577\relax
\mciteBstWouldAddEndPuncttrue
\mciteSetBstMidEndSepPunct{\mcitedefaultmidpunct}
{\mcitedefaultendpunct}{\mcitedefaultseppunct}\relax
\EndOfBibitem
\bibitem[Hachmann \latin{et~al.}(2006)Hachmann, Cardoen, and Chan]{Hachmann}
Hachmann,~J.; Cardoen,~W.; Chan,~G. K.-L. \emph{J. Chem. Phys.} \textbf{2006},
  \emph{125}\relax
\mciteBstWouldAddEndPuncttrue
\mciteSetBstMidEndSepPunct{\mcitedefaultmidpunct}
{\mcitedefaultendpunct}{\mcitedefaultseppunct}\relax
\EndOfBibitem
\bibitem[Fertitta \latin{et~al.}(2014)Fertitta, Paulus, Barcza, and
  Legeza]{paulus}
Fertitta,~E.; Paulus,~B.; Barcza,~G.; Legeza,~O. \emph{Phys. Rev. B}
  \textbf{2014}, \emph{90}, 245129\relax
\mciteBstWouldAddEndPuncttrue
\mciteSetBstMidEndSepPunct{\mcitedefaultmidpunct}
{\mcitedefaultendpunct}{\mcitedefaultseppunct}\relax
\EndOfBibitem
\bibitem[Frisch \latin{et~al.}()Frisch, Trucks, Schlegel, Scuseria, Robb,
  Cheeseman, Scalmani, Barone, Mennucci, Petersson, Nakatsuji, Caricato, Li,
  Hratchian, Izmaylov, Bloino, Zheng, Sonnenberg, Hada, Ehara, Toyota, Fukuda,
  Hasegawa, Ishida, Nakajima, Honda, Kitao, Nakai, Vreven, Montgomery, Peralta,
  Ogliaro, Bearpark, Heyd, Brothers, Kudin, Staroverov, Kobayashi, Normand,
  Raghavachari, Rendell, Burant, Iyengar, Tomasi, Cossi, Rega, Millam, Klene,
  Knox, Cross, Bakken, Adamo, Jaramillo, Gomperts, Stratmann, Yazyev, Austin,
  Cammi, Pomelli, Ochterski, Martin, Morokuma, Zakrzewski, Voth, Salvador,
  Dannenberg, Dapprich, Daniels, Farkas, Foresman, Ortiz, Cioslowski, and
  Fox]{g09}
Frisch,~M.~J.; Trucks,~G.~W.; Schlegel,~H.~B.; Scuseria,~G.~E.; Robb,~M.~A.;
  Cheeseman,~J.~R.; Scalmani,~G.; Barone,~V.; Mennucci,~B.; Petersson,~G.~A.;
  Nakatsuji,~H.; Caricato,~M.; Li,~X.; Hratchian,~H.~P.; Izmaylov,~A.~F.;
  Bloino,~J.; Zheng,~G.; Sonnenberg,~J.~L.; Hada,~M.; Ehara,~M.; Toyota,~K.;
  Fukuda,~R.; Hasegawa,~J.; Ishida,~M.; Nakajima,~T.; Honda,~Y.; Kitao,~O.;
  Nakai,~H.; Vreven,~T.; Montgomery,~J.~A.,~{Jr.}; Peralta,~J.~E.; Ogliaro,~F.;
  Bearpark,~M.; Heyd,~J.~J.; Brothers,~E.; Kudin,~K.~N.; Staroverov,~V.~N.;
  Kobayashi,~R.; Normand,~J.; Raghavachari,~K.; Rendell,~A.; Burant,~J.~C.;
  Iyengar,~S.~S.; Tomasi,~J.; Cossi,~M.; Rega,~N.; Millam,~J.~M.; Klene,~M.;
  Knox,~J.~E.; Cross,~J.~B.; Bakken,~V.; Adamo,~C.; Jaramillo,~J.;
  Gomperts,~R.; Stratmann,~R.~E.; Yazyev,~O.; Austin,~A.~J.; Cammi,~R.;
  Pomelli,~C.; Ochterski,~J.~W.; Martin,~R.~L.; Morokuma,~K.;
  Zakrzewski,~V.~G.; Voth,~G.~A.; Salvador,~P.; Dannenberg,~J.~J.;
  Dapprich,~S.; Daniels,~A.~D.; Farkas,~O.; Foresman,~J.~B.; Ortiz,~J.~V.;
  Cioslowski,~J.; Fox,~D.~J. \textsc{Gaussian 09 {R}evision {C}.01}. Gaussian
  Inc. Wallingford CT 2009\relax
\mciteBstWouldAddEndPuncttrue
\mciteSetBstMidEndSepPunct{\mcitedefaultmidpunct}
{\mcitedefaultendpunct}{\mcitedefaultseppunct}\relax
\EndOfBibitem
\bibitem[Dunning(1989)]{ccbasis}
Dunning,~T.~H. \emph{J. Chem. Phys.} \textbf{1989}, \emph{90}, 1007--1023\relax
\mciteBstWouldAddEndPuncttrue
\mciteSetBstMidEndSepPunct{\mcitedefaultmidpunct}
{\mcitedefaultendpunct}{\mcitedefaultseppunct}\relax
\EndOfBibitem
\bibitem[Kendall \latin{et~al.}(1992)Kendall, Dunning, and
  Harrison]{augccbasis}
Kendall,~R.~A.; Dunning,~T.~H.; Harrison,~R.~J. \emph{J. Chem. Phys.}
  \textbf{1992}, \emph{96}, 6796--6806\relax
\mciteBstWouldAddEndPuncttrue
\mciteSetBstMidEndSepPunct{\mcitedefaultmidpunct}
{\mcitedefaultendpunct}{\mcitedefaultseppunct}\relax
\EndOfBibitem
\end{mcitethebibliography}
\providecommand{\latin}[1]{#1}
\providecommand*\mcitethebibliography{\thebibliography}
\csname @ifundefined\endcsname{endmcitethebibliography}
  {\let\endmcitethebibliography\endthebibliography}{}

\end{document}